\documentclass[prb,showpacs,byrevtex,amsmath,amssymb,preprint]{revtex4}
\usepackage{graphicx}% Include figure files
\usepackage{dcolumn}% Align table columns on decimal point
\usepackage{bm}% bold math
\begin{document}
%\draft
\preprint{PREPRINT}

\title[Short Title]{Behavior of rod-like polyelectrolytes near an oppositely charged surface}
% Force line breaks with \\

%%%%%%%%%%%%%%% AUTHORS %%%%%%%%%%%%%%%%%%%%%%%%%%%
%1st Author
\author{Ren\'e Messina}
%\thanks{Also at Physics Department, XYZ University.}%Lines break automatically or can be forced with \\
\email{messina@thphy.uni-duesseldorf.de}% REVTEX 4
\affiliation
{Institut f\"ur Theoretische Physik II,
Heinrich-Heine-Universit\"at D\"usseldorf,
Universit\"atsstrasse 1,
D-40225 D\"usseldorf,
Germany}

\date{\today}% It is always \today, today, but you may specify any date with

\begin{abstract}
The behavior of highly charged short rod-like polyelectrolytes near oppositely charged 
planar surfaces is investigated by means of Monte Carlo simulations.
A detailed microstructural study, including monomer and fluid charge distribution, 
and chain orientation, is provided. 
The influence of chain length, substrate's surface-charge-density and image forces is considered.
Due to the lower chain-entropy (compared to flexible chains), our simulation data show that 
rod-like polyelectrolytes can, in general, better adsorb than flexible ones do. Nonetheless, 
at low substrate-dielectric-constant, it is found that repulsive image forces tend to significantly 
reduce this discrepancy. 

\end{abstract}
\maketitle

%%%%%%%%%%%%%%%%%%%%%%%%%%%%%%%%%%%%%
\section{Introduction}
%%%%%%%%%%%%%%%%%%%%%%%%%%%%%%%%%%%%%

Polyelectrolyte (PE - i.e., charged polymer) adsorption on charged
surfaces is a versatile process having industrial as well as biological
applications. 
In particular, the case of rod-like (stiff) PEs, which corresponds to the
situation of short DNA fragments or other similar biomaterials, has some relevance
for biological processes such as gene therapy \cite{Raedler_Science_1997}
or biotechnology. 
\cite{Fodor_Science_1997,Pfohl_Langmuir_2001,Pfohl_ChemPhChem_2003}    

From a theoretical viewpoint, the behavior of stiff PEs near an oppositely 
charged surface has been studied by various authors on a mean-field level. 
\cite{Hoagland_Macromol_1990,Menes_EPL_1998,Menes_EPJE_2000,Sens_PRL_2000,
Fleck_PRE_2001,Netz_JPCM_2003,Cheng_JCP_2003}
Menes et al. \cite{Menes_EPJE_2000} considered the interaction between two infinitely 
long charged rods near a salty surface in the framework of the Debye-H\"uckel theory.
Due to the low dimensionality of the system they reported an algebraic decay 
for the effective interaction that was confirmed by their Brownian dynamics simulations. 
\cite{Menes_EPJE_2000}
The more simple and fundamental situation of a \textit{single} and \textit{infinitely long} charged 
rod in the vicinity of a charged surface was investigated by several authors. 
\cite{Menes_EPL_1998,Sens_PRL_2000,Fleck_PRE_2001}
The problem of the so-called ``counterion release'' (i.e., ``Manning decondensation'') 
from a rigid PE approaching an oppositely charged was examined 
by Sens and Joanny \cite{Sens_PRL_2000} and by Fleck and von Gr\"unberg \cite{Fleck_PRE_2001} 
using Poisson-Boltzmann Theory. In a similar spirit, Menes et al. \cite{Menes_EPL_1998}
found that the screening of the adsorbed charged rod's field, due to counterions 
and mobile surface charges, is highly sensitive to the degree of membrane's surface charging.
The adsorption of \textit{short} rod-like PEs was also considered by some researchers. 
\cite{Hoagland_Macromol_1990,Cheng_JCP_2003} 
Recently, Cheng and Olvera de la Cruz \cite{Cheng_JCP_2003} 
investigated the adsorption/desorption transition 
including lateral correlations by assuming a
regular flat lattice for the adsorbed charged rods.
This latter assumption is only valid when the electrostatic rod-surface binding is strong enough.
Closely related to our problem, Hoagland \cite{Hoagland_Macromol_1990} analyzed the monomer 
concentration profile as well as the 
chain-orientation with respect to the charged substrate's surface for a single short rod-like PE. 
The notorious complication of image forces stemming from the dielectric 
discontinuity between the substrate and the solvent was also addressed by some authors.
\cite{Sens_PRL_2000,Netz_JPCM_2003,Netz_Macrom_1999} 
More specifically, for low dielectric constant
(i.e., repulsive image forces) and fixed surface-ions of the substrate:
(i) Sens and Joanny \cite{Sens_PRL_2000} showed that the condensed counterions 
are not always released as the stiff PE approaches the substrate  and 
(ii) Netz \cite{Netz_JPCM_2003} showed an extra decrease of the charge dissociation of the stiff PE 
(i.e., charge regulation in salty solution nearby an uncharged polarized interface).
 
Since those studies \cite{Hoagland_Macromol_1990,Sens_PRL_2000,Netz_JPCM_2003} 
were realized in the framework of the Poisson-Boltzmann theory and for a single chain, 
the relevant phenomenon of \textit{charge reversal} of the substrate's surface-charge by the 
adsorbed PEs can not be captured.

In this paper, we propose to elucidate the microstructural behavior of (very) short 
rod-like PEs near an oppositely charged surface by using Monte Carlo (MC) computer simulations. 
The effect of image forces is also systematically investigated. 
To better understand the effect of chain-entropy, 
a comparison with the previous work of Messina \cite{Messina_PRE_2004} concerning fully 
flexible PEs is carried out.
Our article is organized as follows: The simulation model is detailed in 
Sec. \ref{ Sec.simu_method}. Our results are presented in Sec. \ref{Sec.Results},
and concluding remarks are provided in Sec. \ref{Sec.summary}.  

%%%%%%%%%%%%%%%%%%%%%%%%%%%%%%%%%%%%%
\section{Model and Parameters
\label{ Sec.simu_method}}
%%%%%%%%%%%%%%%%%%%%%%%%%%%%%%%%%%%%%

%%%%%%%%%%%%%%%%%%%%%%%%%%%%%%%%%%%%%
\subsection{Simulation model
\label{ Sec.simu_model}}
%%%%%%%%%%%%%%%%%%%%%%%%%%%%%%%%%%%%%

The model system under consideration is similar to that recently
investigated for the adsorption of flexible chains.
\cite{Messina_macromol_2004,Messina_PRE_2004}
Within the framework of the primitive model we consider a PE
solution near a charged hard wall with an implicit solvent 
of relative permittivity $\epsilon_{solv}\approx 80$
(i.e., water at $z>0$).  
The substrate located at $z<0$ is characterized by  
a relative permittivity $\epsilon_{subs}$ which leads
to a dielectric jump $\Delta_{\epsilon}$ 
(when $\epsilon_{solv} \neq \epsilon_{subs}$) at the interface 
(positioned at $z=0$) defined as 
%
%%%%%%%%%%%%%%%%
\begin{eqnarray}
\label{eq.Delta_eps}
\Delta_{\epsilon}  = \frac{\epsilon_{solv} - \epsilon_{subs}}
                          {\epsilon_{solv} + \epsilon_{subs}}.
\end{eqnarray}
%%%%%%%%%%%%%%%%
%

The \textit{negative} bare surface-charge density of the substrate's interface is $-\sigma_0 e$,
where $e$ is the (positive) elementary charge and $\sigma_0>0$ is the number 
of charges per unit area. 
The latter is always electrically compensated by its accompanying 
monovalent counterions of charge $Z_+e$ 
(i.e., monovalent cations with $Z_{+}=+1$) of diameter $a$.
Rod-like PE chains are made up of $N_m$ \textit{monovalent} positively charged monomers 
($Z_m=Z_+=+1$) of diameter $a$. The bond length $l$ is also set to $l=a$ so that
the length $L_{rod}$ of a rod-like PE is $L_{rod} = N_m l = N_m a$.
The counterions (monovalent anions: $Z_{-}=-1$) of the PEs are also explicitly 
taken into account with the same parameters, up to the charge-sign, as the substrate's
counterions. 
Hence, all the constitutive microions are monovalent ($Z=Z_{+}=Z_m=-Z_{-}=1$) 
and monosized with diameter $a$.
All these particles are immersed in a rectangular $L \times L \times  \tau$ box.
Periodic boundary conditions are applied in the $(x,y)$ directions, 
whereas hard walls are present at $z=0$ (location of the charged interface) 
and $z=\tau$ (location of an \textit{uncharged} wall).
It is to say that we work in the framework of the cell model.

The total energy of interaction of the system can be written as

\begin{eqnarray}
\label{eq.U_tot}
U_{tot} & = &  
\sum_i  \left[ U_{hs}^{(intf)}(z_i) + U_{coul}^{(intf)}(z_i) \right] + 
\\ \nonumber
&& \sum _{i,i<j} \left[U_{hs}^{(mic)}(r_{ij}) + U_{coul}^{(mic)}({\bf r}_i, {\bf r}_j)\right],
\end{eqnarray}
where the first (single) sum stems from the interaction between a microion $i$ 
[located at $z=z_i$ with $i=(+,-,m)$] and the charged interface, 
and the second (double) sum stems from the pair interaction between microions 
$i$ and $j$ with $r_{ij}=|{\bf r}_i - {\bf r}_j|$.
All these contributions to $U_{tot}$ in Eq. (\ref{eq.U_tot})
are described in detail below.

Excluded volume interactions are modeled via a hardcore potential 
defined as follows
%
%%%%%%%%%%%%%%%%%%%%%%%%%%%%%%%%%%%%%%%%%
\begin{equation}
\label{eq.U_hs}
U_{hs}^{(mic)}(r_{ij})=\left\{
\begin{array}{ll}
0,
& \mathrm{for}~r_{ij} \geq a \\
\infty,
& \mathrm{for}~r_{ij} < a 
\end{array}
\right.
\end{equation}
%%%%%%%%%%%%%%%%%%%%%%%%%%%%%%%%%%%%%%%%%
%
for the microion-microion one, and
%
%%%%%%%%%%%%%%%%%%%%%%%%%%%%%%%%%%%%%%%%%
\begin{equation}
\label{eq.U_hs_plate}
U_{hs}^{(intf)}(z_i)=\left\{
\begin{array}{ll}
0,
& \mathrm{for} \quad a/2 \leq z_i \leq  \tau - a/2 \\
\infty,
& \mathrm{otherwise}
\end{array}
\right.
\end{equation}
%%%%%%%%%%%%%%%%%%%%%%%%%%%%%%%%%%%%%%%%%
%
for the interface-microion one.

The electrostatic energy of interaction between two microions $i$ and $j$ reads
%
%%%%%%%%%%%%%%%%%%%%%%%%%%%%%%%%%
\begin{equation}
\label{eq.U_coul} 
\beta U_{coul}^{(mic)}({\bf r}_i, {\bf r}_j) =
Z_i Z_j l_B \left[  
  \frac{1}{r_{ij}} 
+ \frac{\Delta_{\epsilon}}{\sqrt{x_{ij}^2 + y_{ij}^2 + (z_i+z_j)^2}}
\right],
\end{equation}
%%%%%%%%%%%%%%%%%%%%%%%%%%%%%%%%%
%
where $l_{B}=\beta e^{2}/(4\pi \epsilon _{0}\epsilon _{solv})$ is the Bjerrum
length corresponding to the distance at which two protonic charges
interact with $1/\beta=k_B T$, and $\Delta_{\epsilon}$ is given by 
Eq.~\eqref{eq.Delta_eps}.  
The first term in Eq.~\eqref{eq.U_coul} corresponds
to the direct Coulomb interaction between real microions,
whereas the second term represents the interaction between the real microion $i$
and the image of microion $j$. By symmetry, the latter also describes 
the interaction between the real microion $j$ and the image of microion $i$ 
yielding an implicit prefactor $1/2$ in Eq.\eqref{eq.U_coul}.
The electrostatic energy of interaction between a microion $i$ and the
(uniformly) charged interface reads
%
%%%%%%%%%%%%%%%%%%%%%%%%%%%%%%%%%
\begin{equation}
\label{eq.U_coul_plate} 
\beta U_{coul}^{(intf)}(z_i) =
l_B \left[ 
2 \pi  Z_i (1+\Delta_{\epsilon})\sigma_0 z_i
+ \frac{Z_i^2\Delta_{\epsilon}}{4z_i}
\right].
\end{equation}
%%%%%%%%%%%%%%%%%%%%%%%%%%%%%%%%%
%
The second term in Eq.\eqref{eq.U_coul_plate} stands for the \textit{self-image} 
interaction, i.e., the interaction between the microion $i$ and its own image.
An appropriate and efficient modified Lekner sum was utilized to compute 
the electrostatic interactions with periodicity in \textit{two} 
directions. \cite{Brodka_MolPhys_2002}
This latter technique was already successfully applied to the case of PE
multilayering \cite{Messina_macromol_2004} and polycation adsorption. 
\cite{Messina_PRE_2004}
To link our simulation parameters
to experimental units and room temperature ($T=298$K) we choose
$a =4.25$ \AA\ leading to the Bjerrum length of water
$l_{B}=1.68a =7.14$ \AA. In order to investigate  the effect of
image forces we take a value of $\epsilon_{subs}=2$ for the
dielectric constant of the charged substrate 
(which is a typical value for silica or mica substrates \cite{Malinsky_JPCB_2001}) 
and $\epsilon_{solv}=80$ for that of the aqueous solvent yielding
$\Delta_{\epsilon}=\frac{80 - 2}{80+2} \approx 0.951$.
The case of identical dielectric constants $\epsilon_{subs}=\epsilon_{solv}$
($\Delta_{\epsilon}=0$) corresponds to the situation where there are no
image charges.

%
%%%%%%%%%%%%%%%%%%%%%%%%%%%%%%%%%%%%%%%%%%%%%%%%%%%%%%%%
% TABLE 1
\begin{table}[b]
\caption{
List of key parameters with some fixed values.
}
\label{tab.simu-param}
\begin{ruledtabular}
\begin{tabular}{lc}
 Parameters&
\\
\hline
 $T=298K$&
 room temperature\\
 $\sigma_0 L^2$&
 charge number of the substrate\\
 $\Delta_{\epsilon} = 0 ~ {\rm or} ~ 0.951$ &
 dielectric discontinuity\\
 $Z=1$&
 microion valence\\
 $a =4.25$ \AA\ &
 microion diameter\\
 $l_{B}=1.68a =7.14$ \AA\ &
 Bjerrum length\\
 $L=25 a $&
 $(x,y)$-box length\\
 $\tau=75 a $&
 $z$-box length\\
 $N_{rod}$&
 number of rod-like PEs\\
 $N_m$&
 number of monomers per rod-like chain
\end{tabular}
\end{ruledtabular}
\end{table}
%%%%%%%%%%%%%%%%%%%%%%%%%%%%%%%%%%%%%%%%%%%%%%%%%%%%%%%%
%

All the simulation parameters are gathered in Table \ref{tab.simu-param}.
The set of simulated systems can be found in Table \ref{tab.simu-runs}.
The equilibrium properties of our model system were obtained by 
using standard canonical MC simulations following the Metropolis scheme. 
\cite{Metropolis_JCP_1953,Allen_book_1987}
In detail:
%%%%%%%%%%%%%%%
\begin{itemize}
%cou 
\item Single particle (translational) moves were applied to the counterions 
      (i.e., anions and cations) with an acceptance ratio of $50\%$.
%rod  
\item As far as trial moves for the rod-like PEs are concerned and given the 
      anisotropy of these objects, random translational moves as well as 
      \textit{rotational} ones were performed at the same frequency. \cite{Blaak_JCP_1999}
      Random rotational moves were achieved by choosing randomly new  
      orientation-vectors of the rod-like particles.
      This method is 
      (i) computationally not too demanding, 
      (ii) leads to an efficient configurational space sampling and 
      (iii) fulfills the rules of detailed-balance. 
      The acceptance ratio was also set to $50\%$.
\end{itemize}
%%%%%%%%%%%%%%%
The total length of a simulation run is set to $3\times 10^6$ MC steps per particle.
Typically, about $10^5$ MC steps were required for equilibration, and 
$2.5 \times 10^6$ MC steps were used to perform measurements. 
%\cite{note_barrier}.

%%%%%%%%%%%%%%%%%%%%%%%%%%%%%%%%%%%%%%%%%%%%%%%%%%%%%%%%
% TABLE 2
\begin{table}
\caption{
Simulated systems' parameters. The number of counterions (cations and anions) ensuring
the overall electroneutrality of the system is not indicated.
}
\label{tab.simu-runs}
\begin{ruledtabular}
\begin{tabular}{lccc}
 System&
 $N_{rod}$&
 $N_m$&
 $\sigma_0L^2$
\\
\hline
 $A$&
 $96$&
 $2$&
 $64$\\
 $B$&
 $48$&
 $4$&
 $64$\\
 $C$&
 $32$&
 $6$&
 $64$\\
 $D$&
 $24$&
 $8$&
 $64$\\
 $E$&
 $16$&
 $12$&
 $64$\\
 $F$&
 $24$&
 $8$&
 $32$\\
$G$&
 $24$&
 $8$&
 $128$
\end{tabular}
\end{ruledtabular}
\end{table}
%%%%%%%%%%%%%%%%%%%%%%%%%%%%%%%%%%%%%%%%%%%%%%%%%%%%%%%%

%%%%%%%%%%%%%%%%%%%%%%%%%%%%%%%%%%%%%
\subsection{Measured quantities
 \label{Sec.Target}}
%%%%%%%%%%%%%%%%%%%%%%%%%%%%%%%%%%%%%

We briefly describe the different observables that are going to be measured.  
In order to study the PE adsorption, we compute the monomer density
$n(z)$ that is normalized as follows

%%%%%%%%%%%%%%%%%%%%
\begin{equation}
\label{eq.n_z}
\int ^{\tau-a/2}_{a/2} n(z) L^2 dz = N_{rod} N_m.
\end{equation}
%%%%%%%%%%%%%%%%%%%%
%
%
To further characterize the PE adsorption, we also compute the total number of  
accumulated monomers $\bar{N}(z)$ within a distance $z$
from the charged interface that is given by
%
%%%%%%%%%%%%%%%%%%%%
\begin{equation}
\label{eq.N_z}
\bar{N}(z) = \int ^{z}_{a/2} n(z') L^2 dz'.
\end{equation}
%%%%%%%%%%%%%%%%%%%%
%
It is useful to introduce the fraction of adsorbed monomers,
$N^*(z)$, which is defined as follows
%
%%%%%%%%%%%%%%%%%%%%
\begin{equation}
\label{eq.N_z_star}
N^*(z) = \frac{\bar{N}(z)}{N_{rod} N_m}.
\end{equation}
%%%%%%%%%%%%%%%%%%%%
%

The \textit{orientation} of the rod-like PEs can be best monitored
by the angle $\theta$ formed between the $z$-axis and 
the PE-axis. \cite{Hoagland_Macromol_1990}
A convenient quantity is provided by its second order Legendre polynomial:
%
%%%%%%%%%%%%%%%%%%%%
\begin{equation}
\label{eq.legendre_pol}
 P_2 \left[ \cos \theta(z) \right] 
 = \frac{1}{2} \left[3 \cos^2 \theta(z) -1 \right],
\end{equation}
%%%%%%%%%%%%%%%%%%%%
%
where $z$ corresponds to the smallest wall-monomer distance for a given PE.
Thereby
%
%%%%%%%%%%%%%%%%%%%%
\begin{equation}
\label{eq.Sz}
S(z) \equiv  \langle P_2 \left[ \cos \theta(z) \right] \rangle
\end{equation}
%%%%%%%%%%%%%%%%%%%%
%
takes the values $-\frac{1}{2}$, 0, and +1 for PEs 
that are perpendicular, randomly oriented, and parallel to the 
$z$-axis, respectively.

Another relevant quantity is the global \textit{net fluid
charge} $\sigma(z)$ which reads
%%%%%%%%%%%%%%%%%%%%%%%%%%%%%%%%%%%%%%%%%%%%%%%%%%%%%
\begin{equation}
\label{Eq.Qz}
\sigma(z)=\int ^{z}_{a/2} \left[
n_{+}(z') - n_{-}(z')\right] dz',
\end{equation}
%%%%%%%%%%%%%%%%%%%%%%%%%%%%%%%%%%%%%%%%%%%%%%%%%%%%%
%
where $n_+$ and $n_-$ stand for the density of all the
positive microions (i.e., monomers and substrate's counterions) and
negative microions (i.e., PEs' counterions), respectively.
The corresponding reduced surface charge density
$\sigma^*(z)$ is given by:
%%%%%%%%%%%%%%%%%%%%%%%%%%%%%%%%%%%%%%%%%%%%%%%%%%%%%
\begin{equation}
\label{Eq.Qz_star}
\sigma^*(z) = \frac{\sigma(z)}{\sigma_0}.
\end{equation}
%%%%%%%%%%%%%%%%%%%%%%%%%%%%%%%%%%%%%%%%%%%%%%%%%%%%%
%
Thereby, $\sigma^*(z)$ corresponds, up to a prefactor $\sigma_0 e$, 
to the net fluid charge per unit area 
(omitting the surface charge density $-\sigma_0e$ of the substrate)
within a distance $z$ from the charged wall. 
At the uncharged wall, electroneutrality imposes $\sigma^*(z=\tau-a/2)=1$.  
By simple application of the Gauss' law,
$\left[ \sigma^*(z)-1\right]$ is directly proportional
to the mean electric field at $z$.  Therefore $\sigma^*(z)$ can
measure the \textit{screening} strength of the substrate
by the neighboring solute charged species.

%%%%%%%%%%%%%%%%%%%%%%%%%%%%%%%%%%%%%%%%%%
\section{Results and discussion
 \label{Sec.Results}}
%%%%%%%%%%%%%%%%%%%%%%%%%%%%%%%%%%%%%%%%%%

It is well known that the effects of image forces become especially
relevant at low surface charge density of the interface. 
Furthermore, it is also clear that the self-image interaction 
(\textit{repulsive} for $\Delta_{\epsilon}>0$, as is presently the case)
is higher the higher the charge of the ions (polyions). 
Hence, we are going to study (i) the influence of chain length 
(Sec. \ref{Sec.chain_length}) and (ii) that of surface charge density
(Sec. \ref{Sec.charge_density}).
For the sake of consistency, we set the total number of monomers
to $N_{rod}N_m=192$ meaning that the monomer concentration is 
\textit{fixed} leading to a PE volume fraction 
$\phi = \frac{4\pi}{3}\frac{N_{rod}N_m(a/2)^3}{L^2\tau}\approx 2.14 \times 10^{-3}$ 
(see also Table \ref{tab.simu-runs}).

%%%%%%%%%%%%%%%%%%%%%%%%%%%%%%%%%%%%%
\subsection{Influence of chain length
 \label{Sec.chain_length}}
%%%%%%%%%%%%%%%%%%%%%%%%%%%%%%%%%%%%%

In this part, we consider the influence of chain length $N_m$
at fixed surface-charge-density parameter $\sigma_0L^2=64$.
The latter would experimentally  correspond to a moderate \cite{Tulpar_JPCB_2004}
surface charge density with $-\sigma_0e \approx -0.091 ~ {\rm C/m^2}$.
The chain length is varied from $N_m=2$ up to $N_m=12$ 
(systems $A-E$, see Table \ref{tab.simu-runs}).
We have ensured that, for the longest chains with $N_m=12$,
finite size effects are not important since there 
$L_{rod}=12a$ which is significantly smaller than $L=25a$ or $\tau=75a$.

%%%%%%%%%%%%%%%%%%%%%%%%%%%%%%%%%%%%%
\subsubsection{Monomer distribution
 \label{Sec.mon_dist_Qp-64}}
%%%%%%%%%%%%%%%%%%%%%%%%%%%%%%%%%%%%%

%%%%%%%%%%%%%%%%%%%%%%%%%%%%%%%%%%%%%%%%
% FIG 1
\begin{figure}[b]
\includegraphics[width = 8.0 cm]{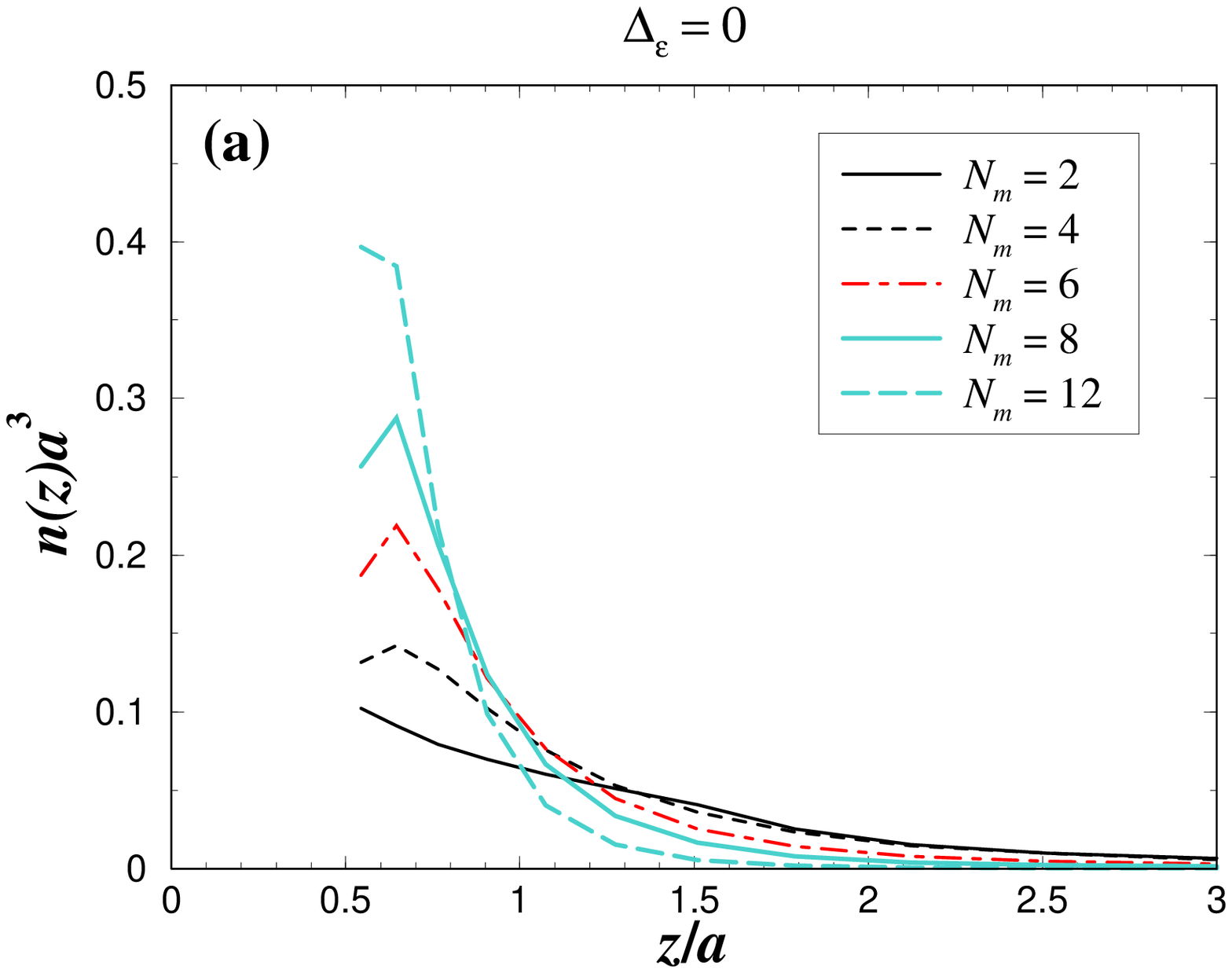}
\includegraphics[width = 8.0 cm]{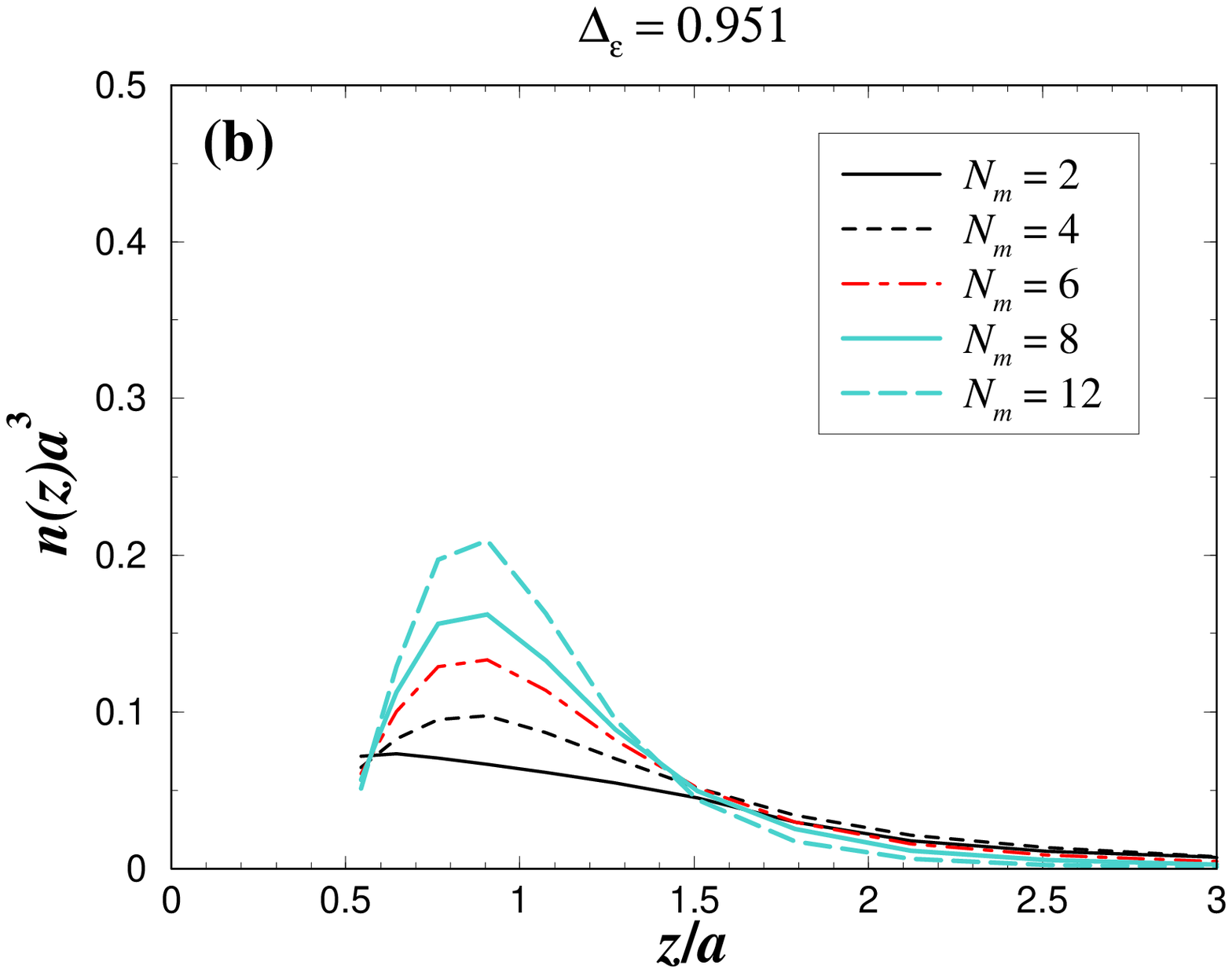}
\caption{
Profiles of the monomer density $n(z)$ for different chain length $N_m$ 
with $\sigma_0L^2=64$ (systems $A-E$).
(a) $\Delta_{\epsilon}=0$.
(b) $\Delta_{\epsilon}=0.951$.
}
\label{fig.nz_Qp-64}
\end{figure}
%%%%%%%%%%%%%%%%%%%%%%%%%%%%%%%%%%%%%%%%
%

%%%%%%%%%%%%%%%%%%%%%%%%%%%%%%%%%%%%%%%%
% FIG 2
\begin{figure}
\includegraphics[width = 8.0 cm]{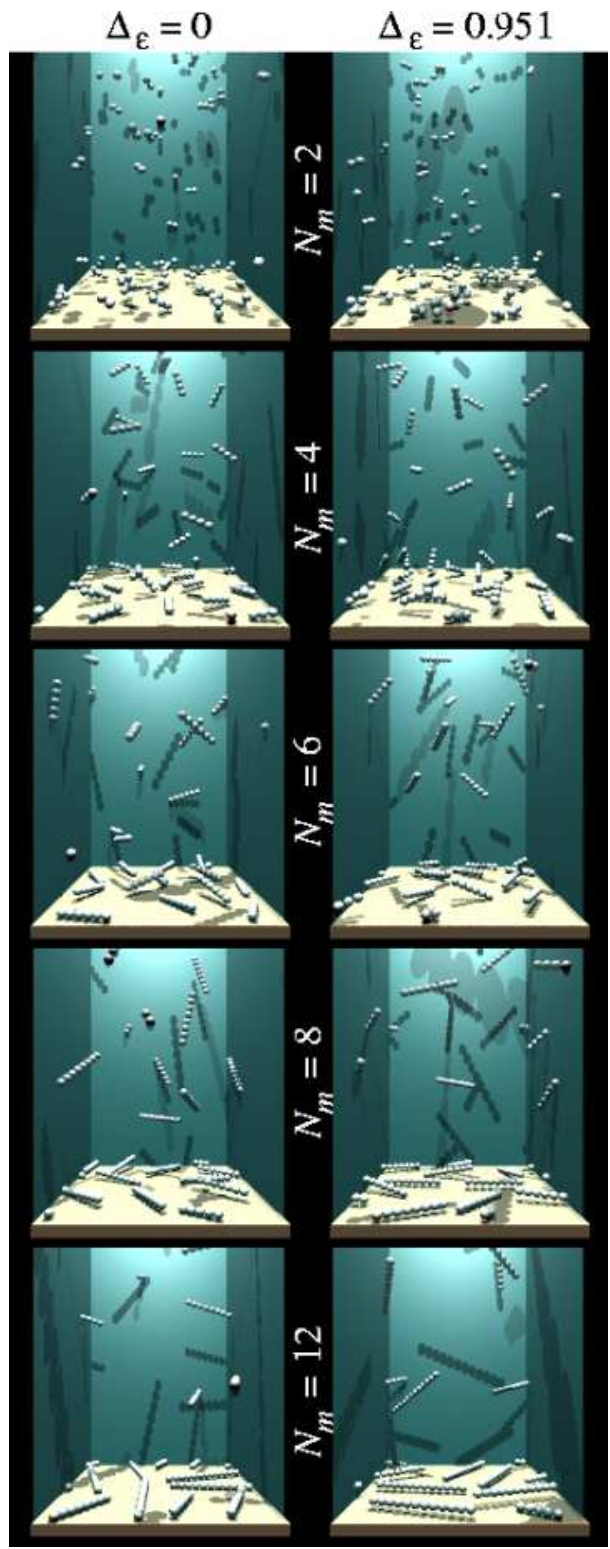}
\caption{Typical equilibrium microstructures of systems $A-E$.
The little counterions are omitted for clarity.}
\label{fig.snap_Qp-64}
\end{figure}
%%%%%%%%%%%%%%%%%%%%%%%%%%%%%%%%%%%%%%%%
%
The profiles of the monomer distribution $n(z)$ can be found in
Fig. \ref{fig.nz_Qp-64} and the corresponding microstructures are sketched in 
Fig. \ref{fig.snap_Qp-64}. 
When no image charges are present [$\Delta_{\epsilon}=0$ - Fig. \ref{fig.nz_Qp-64}(a)],
the monomer-density profile $n(z)$ exhibits a monotonic behavior for very
short rigid chains (here $N_m=2$).
For longer chains there exits a small monomer depletion near the charged wall for an 
\textit{intermediate} regime of $N_m$ (here $4 \leq N_m \leq 8$). At high enough enough $N_m$ 
(here $N_m=12$) our simulation data reveal again a monotonic behavior of $n(z)$.
This interesting behavior is the results of two antagonistic driving forces, namely (i) 
chain-entropy and (ii) the electrostatic wall-monomer attraction.
More precisely, the mechanisms responsible for this $N_m$-induced reentrant behavior 
at $\Delta_{\epsilon}=0$ are as follows:
% reentrance
%%%%%%%%%%%%%%
\begin{itemize}
\item For very short chains (here $N_m=2$) chain-entropy effects are negligible 
      so that one gets a similar behavior to that of point-like counterions 
      with the usual monotonic decaying $n(z)$-profile.
\item The chain-entropy loss (per chain) by adsorption should typically scale like 
      $\ln N_m$ whereas the electrostatic wall-chain attraction scales like $N_m$ explaining 
      why at high enough $N_m$ a purely \textit{effective} attractive wall-monomer interaction 
      is recovered.  
\end{itemize}
%%%%%%%%%%%%%%

Upon polarizing the charged interface  
[$\Delta_{\epsilon}=0.951$ - Fig. \ref{fig.nz_Qp-64}(b)]
the PE adsorption becomes weaker and the $n(z)$-profile
more broadened due to the repulsive image-polyion interaction. 
For $N_m \leq 4$, $n(z)$ presents a maximum at $z=z^*\approx 0.9a$ 
that can be seen as the \textit{thickness} of the adsorbed PE layer.
%by $z^*$ increases with $N_m$.
%This phenomenon is of course due to the fact that the image-polyion repulsion 
%increases with growing $N_m$, similarly to what happens with multivalent (point-like or spherical) 
%counterions. \cite{Torrie_JCP_1982,Messina_image_2002}
Interestingly, the monomer density at contact \textit{decreases} 
with increasing $N_m$. 
This is the result of a \textit{combined} effect of (i) chain-entropy loss 
near the interface and (ii) the $N_m$-induced image-polyion repulsion. 
All those features are well illustrated on the microstructures 
of Fig. \ref{fig.snap_Qp-64}.

\begin{figure}
\includegraphics[width = 8.0 cm]{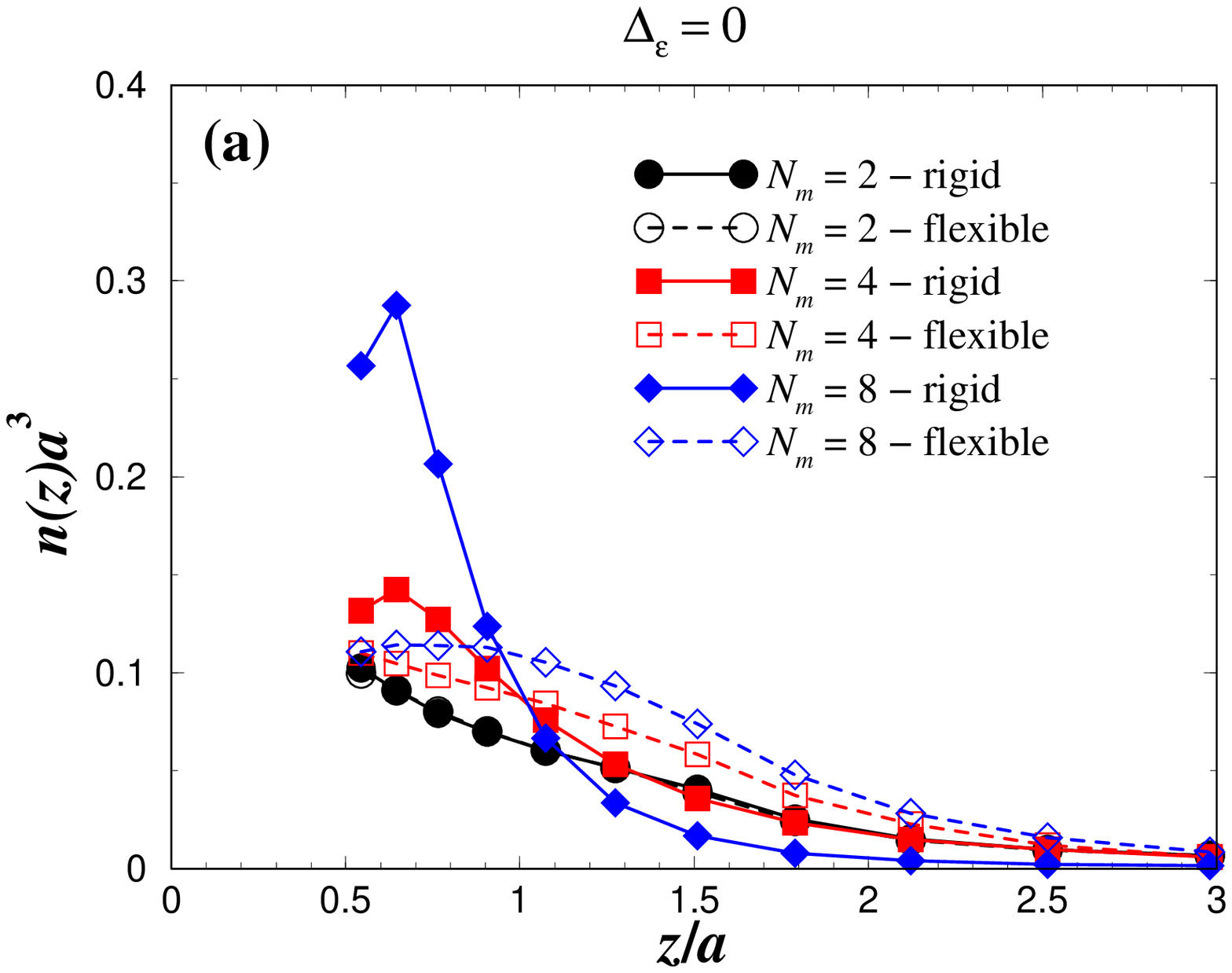}
\includegraphics[width = 8.0 cm]{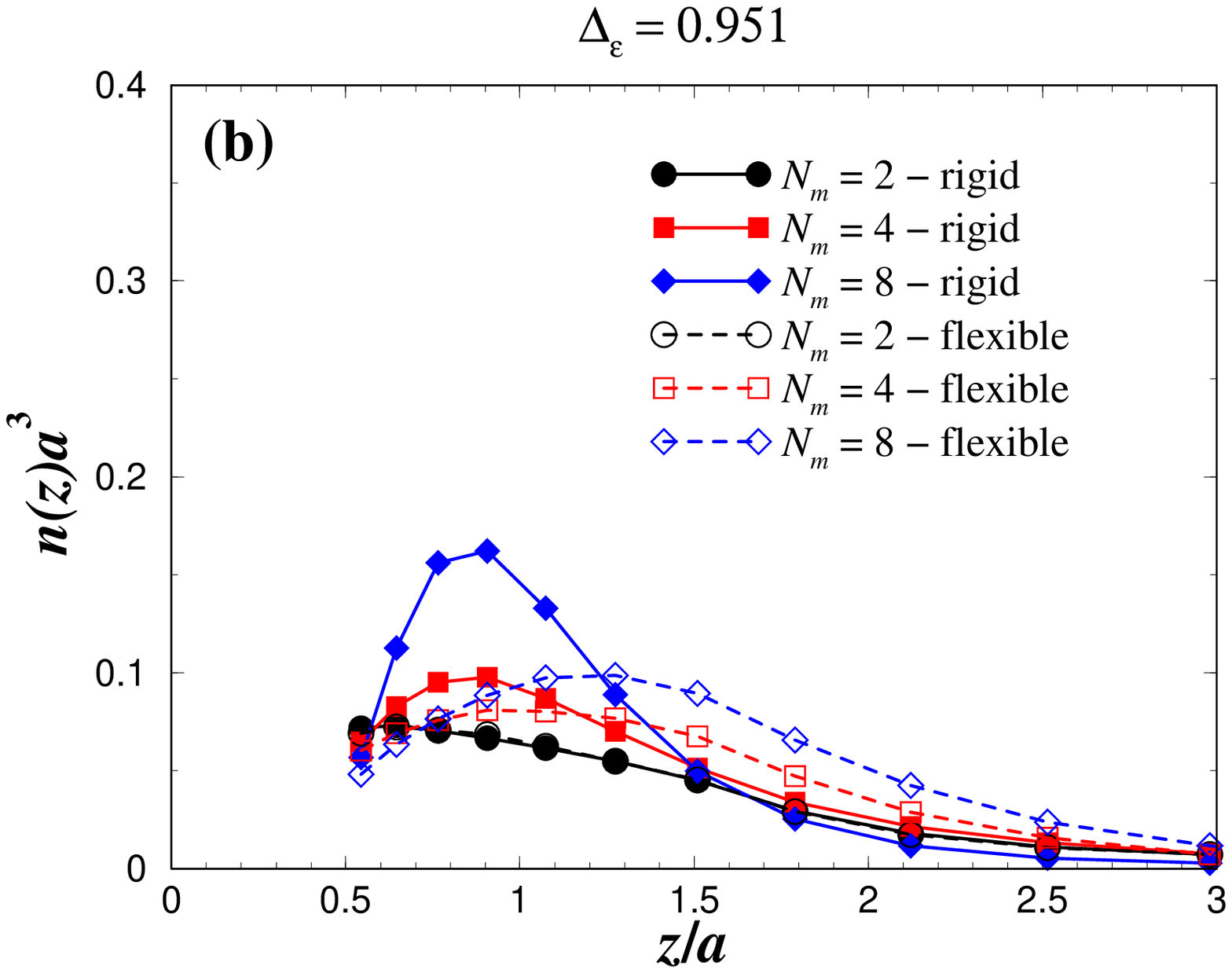}
\caption{
Comparison between \textit{flexible} and rod-like (rigid) PEs for the profiles 
of the monomer density $n(z)$ at different chain length $N_m$. 
(a) $\Delta_{\epsilon}=0$.
(b) $\Delta_{\epsilon}=0.951$.
}
\label{fig.nz_ROD_VS_FLEXIBLE_Qp-64}
\end{figure}

It is instructive to compare the above findings with those obtained 
for \textit{fully flexible} chains.
To do so, we use existing MC data for flexible chains from our previous work 
\cite{Messina_PRE_2004} where all the parameters, up to the chain flexibility, 
are identical with those presently employed for rod-like PEs. 
The comparison is provided in Fig. \ref{fig.nz_ROD_VS_FLEXIBLE_Qp-64}.  
At  $\Delta_{\epsilon}=0$ [see Fig. \ref{fig.nz_ROD_VS_FLEXIBLE_Qp-64}(a)], 
the $n(z)$-profiles for flexible and rigid PEs are quasi-identical 
for $N_m=2$ as it should be. 
For longer chains ($N_m=8$), we clearly see at  $\Delta_{\epsilon}=0$ 
that the degree of adsorption as indicated by the value of $n(z)$ near contact 
is considerably stronger for rigid chains. 
This feature is due to entropy and electrostatic effects. 
Indeed, in the bulk and at given degree of polymerization $N_m$, the chain-entropy 
associated to rigid PEs is much lower than that associated to flexible chains, so that
chain-entropy loss upon adsorption is reduced for rigid chains. 
Secondly and concomitantly, the wall-PE attraction is more efficient for rigid chains 
than for flexible chains because in the latter case the $z$-\textit{fluctuations} of 
the charged monomers are more important.

%%%%%%%%%%%%%%%%%%%%%%%%%%%%%%%%%%%%%%%%
% FIG 3
%%%%%%%%%%%%%%%%%%%%%%%%%%%%%%%%%%%%%%%%
%

As far as the monomer density at true contact is concerned 
[i.e., $n(z \to a/2)$], it seems that, for flexible PEs,
its value is nearly independent of $N_m$ as already 
reported in Ref. \cite{Messina_PRE_2004}.
For a single ionic species of \textit{spherical} shape, a variant
of the contact theorem provides the exact relation:
$n(a/2)-n(\tau-a/2)=2\pi\sigma_0^2 l_B$ yielding
to $n(a/2) \approx 0.11a^{-3}$, which is surprisingly in remarkable agreement with
the value reported in Fig. \ref{fig.nz_ROD_VS_FLEXIBLE_Qp-64}(a) for flexible PE
(and rigid ones for $N_m\leq 4$).
Nonetheless,  Fig. \ref{fig.nz_ROD_VS_FLEXIBLE_Qp-64}(a) shows, already with $N_m=8$, 
a strong deviation from the contact theorem 
(which in principle only holds for structureless spherical ions) 
for rigid PEs, as expected.

The scenario becomes qualitatively different when $\Delta_{\epsilon}=0.951$ 
[see Fig. \ref{fig.nz_ROD_VS_FLEXIBLE_Qp-64}(b)], 
where the $n(z)$-profiles for flexible and rigid PEs become more similar. 
It is to say that the image-polyion repulsion tends to cancel chain-entropy 
effects. 
A closer look at Fig. \ref{fig.nz_ROD_VS_FLEXIBLE_Qp-64}(b) reveals however
that, at given $N_m$, the degree of PE adsorption is 
systematically larger for rigid PEs than for flexible ones as expected.
Those relevant findings can be summarized as follows:
%
%ROD VS FLEXIBLE SUMMARY
%%%%%%%%%%%%%%%
\begin{itemize}
\item Without dielectric discontinuity ($\Delta_{\epsilon}=0$) rigid PE chains can much better 
      adsorb than flexible ones at oppositely charged surfaces essentially because of their 
      significant lower chain-entropy.
\item In the presence of polarization charges ($\Delta_{\epsilon}=0.951$) the degree of PE adsorption
      becomes significantly less sensitive to the chain stiffness. 
\end{itemize}
%%%%%%%%%%%%%%%

%%%%%%%%%%%%%%%%%%%%%%%%%%%%%%%%%%%%%%%%
% FIG 4
\begin{figure}
\includegraphics[width = 8.0 cm]{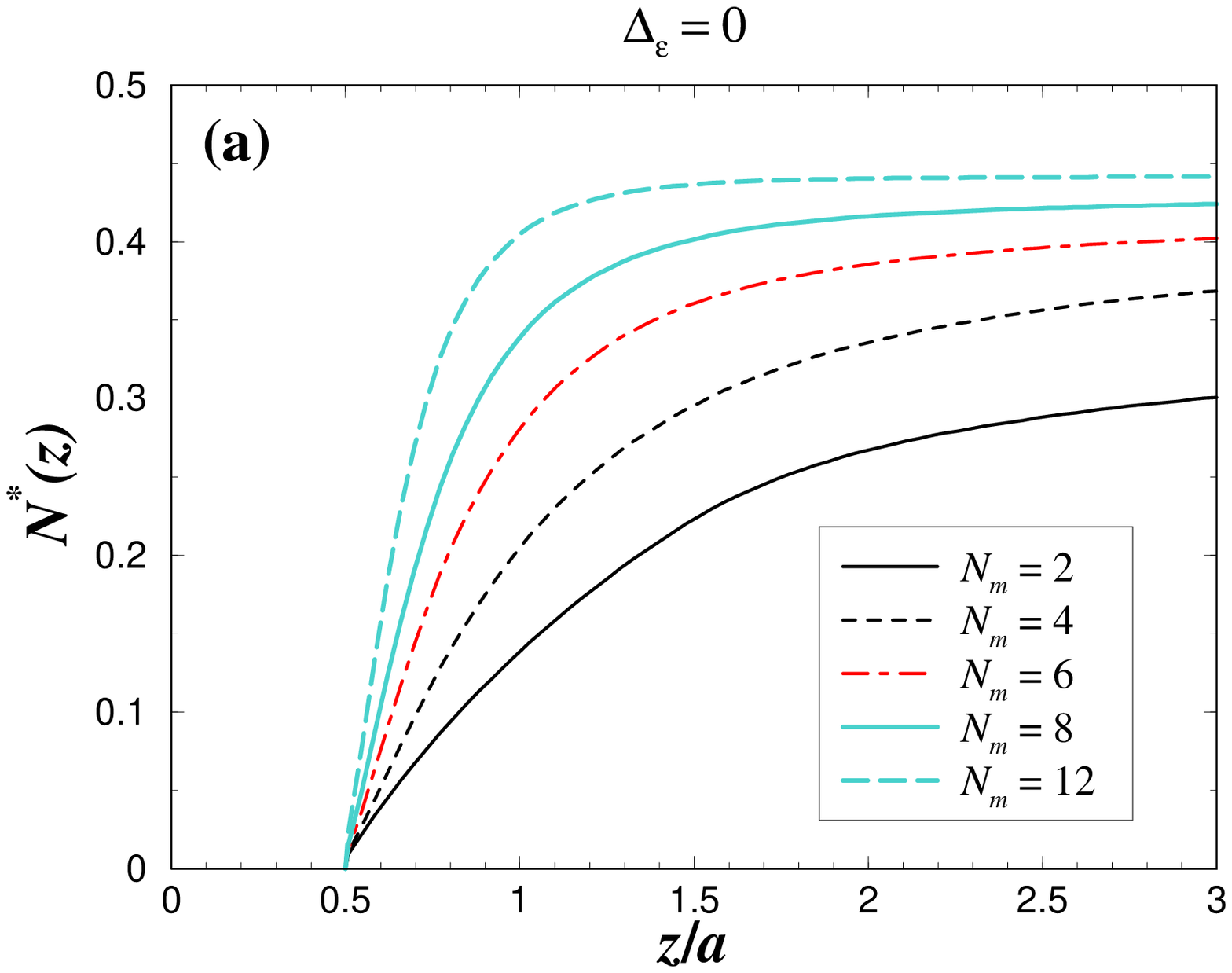}
\includegraphics[width = 8.0 cm]{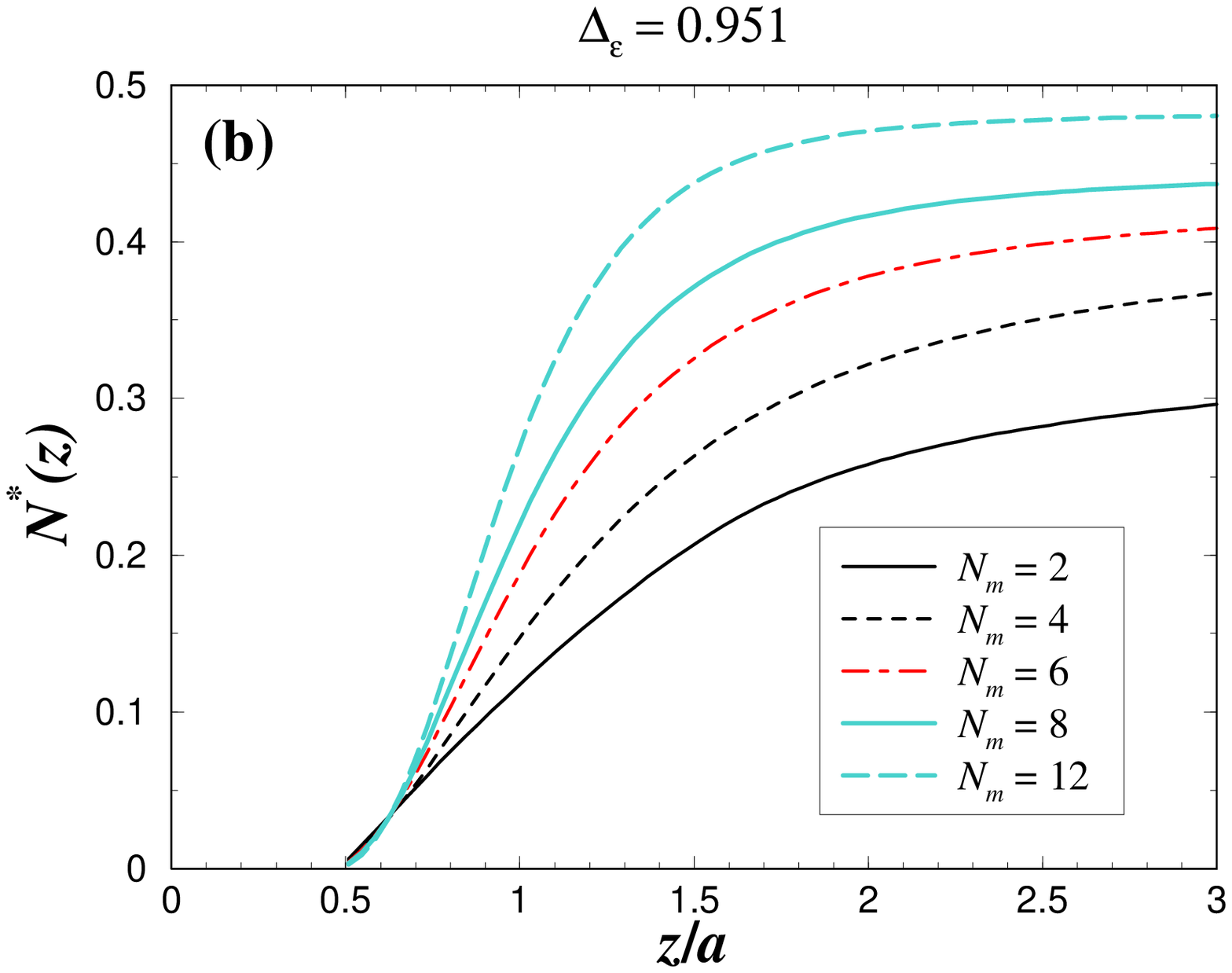}
\caption{
Profiles of the fraction of adsorbed monomers $N^*(z)$ 
for different chain length $N_m$ with $\sigma_0L^2=64$ (systems $A-E$).
(a) $\Delta_{\epsilon}=0$.
(b) $\Delta_{\epsilon}=0.951$. The inset is merely a magnification near contact.
}
\label{fig.Nz_star_Qp-64}
\end{figure}
%%%%%%%%%%%%%%%%%%%%%%%%%%%%%%%%%%%%%%%%
%

In order to quantify the amount of adsorbed monomers as a function of 
the distance $z$ from the charged wall, 
we have also studied $N^*(z)$ as defined by Eq. \eqref{eq.N_z_star}.
Our results are reported in Fig. \ref{fig.Nz_star_Qp-64}.
At $\Delta_{\epsilon}=0$ [see Fig. \ref{fig.Nz_star_Qp-64}(a)] 
the fraction of adsorbed monomers $N^*(z)$ is always larger with growing $N_m$ 
even near the interface. 
On the other hand, at $\Delta_{\epsilon}=0.951$ [see Fig. \ref{fig.Nz_star_Qp-64}(b)]:
(i) $N^*(z)$ gets smaller with growing $N_m$ near the interface 
(roughly for $z/a \lesssim 0.65$) and (ii) $N^*(z)$ is considerably
reduced compared to the $\Delta_{\epsilon}=0$-case. 
For instance (with $N_m=12$) at $z/a=0.9$ 
(corresponding to a layer thickness at $\Delta_{\epsilon}=0.951$), 
$N^*(z)$  can be as large as 0.4 for $\Delta_{\epsilon}=0$ 
against only 0.2 for $\Delta_{\epsilon}=0.951$.

%%%%%%%%%%%%%%%%%%%%%%%%%%%%%%%%%%%%%
\subsubsection{PE orientation
 \label{Sec.legendre_Qp-64}}
%%%%%%%%%%%%%%%%%%%%%%%%%%%%%%%%%%%%%

\begin{figure}
\includegraphics[width = 8.0 cm]{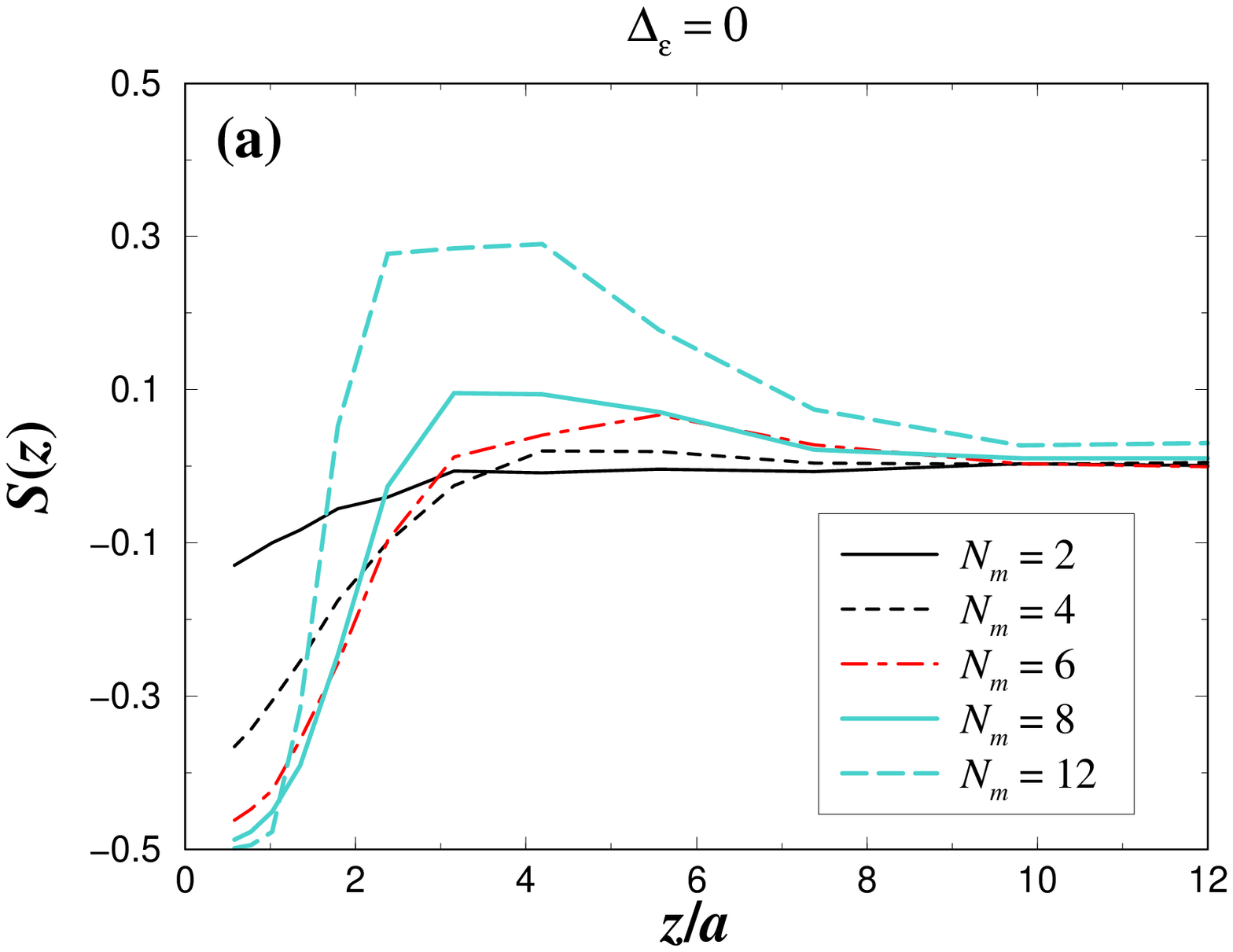}
\includegraphics[width = 8.0 cm]{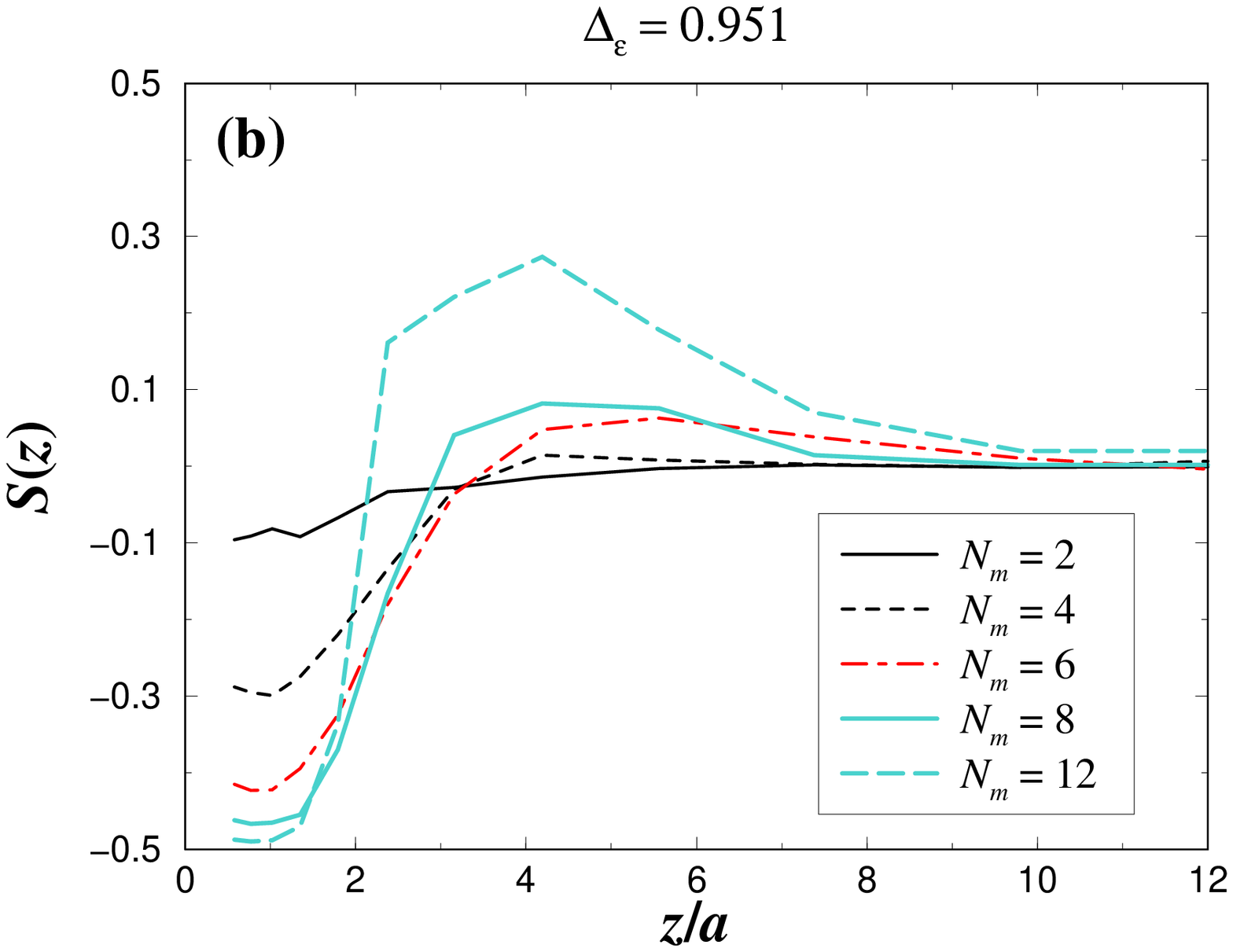}
\caption{
Profiles of $S(z)$ for different chain length $N_m$ 
with $\sigma_0L^2=64$ (systems $A-E$).
(a) $\Delta_{\epsilon}=0$.
(b) $\Delta_{\epsilon}=0.951$.
}
\label{fig.Sz_Qp-64}
\end{figure}

To gain further insight into the properties of rod-like PE adsorption, we
have plotted $S(z)$ as given by Eq.\eqref{eq.Sz} in Fig. \ref{fig.Sz_Qp-64}
so as to characterize the PE orientation with respect to the charged interface.
At $\Delta_{\epsilon}=0$, Fig. \ref{fig.Sz_Qp-64}(a) shows that in the vicinity 
of the interface (roughly for $z \lesssim a$) that the rod-like PEs tend 
to be parallel to the interface-plane with growing $N_m$, 
i.e., $S(z) \rightarrow -1/2$  (see also Fig. \ref{fig.snap_Qp-64}). 
This effect is obviously due to the electrostatic wall-PE binding whose strength 
increases linearly with $N_m$. 
For the longest chains ($N_m=12$)  non-negligible \textit{positive} $S(z)$-values are reported
at intermediate distance from the wall (roughly for $2 \lesssim z/a \lesssim 6$)
signaling a non-trivial orientation correlation with respect to the interface-plane
that will be properly discussed later.
Sufficiently away from the wall, the rod-like PEs are randomly oriented 
[i. e., $S(z) \rightarrow 0$] as it should be.

%%%%%%%%%%%%%%%%%%%%%%%%%%%%%%%%%%%%%%%%
% FIG 5
%%%%%%%%%%%%%%%%%%%%%%%%%%%%%%%%%%%%%%%%
%

In the presence of image forces [$\Delta_{\epsilon}=0.951$ - see Fig. \ref{fig.Sz_Qp-64}(b)] 
the $S(z)$ behavior is more complex. 
A comparison with Fig. \ref{fig.Sz_Qp-64}(a) corresponding to $\Delta_{\epsilon}=0$
immediately shows that repulsive image forces tend to inhibit the alignment of the rod-like PEs 
with respect to the interface-plane near contact. This effect will be  especially vivid
at lower surface charge density $\sigma_0$, as we are going to show later.
%Besides of that, repulsive image forces \textit{also} destabilize the degree of wall-PE
%alignement \textit{within} the PE layer leading to a \textit{non-monotonic} behavior of $S(z)$ near contact
%[see Fig. \ref{fig.Sz_Qp-64}(b)]. 

The non-monotonic behavior of $S(z)$ near contact at $\Delta_{\epsilon}=0.951$, 
similar to that reported for $n(z)$ in Fig. \ref{fig.nz_Qp-64}(b), 
is the result of two antagonistic forces: 
(i) the repulsive image driving force that scales like $1/z$ and 
(ii) the attractive  wall-monomer one that scales like $z$.
As in the case with $\Delta_{\epsilon}=0$, 
(i) the degree of PE-wall parallelism increases with growing $N_m$ near contact and 
(ii) far enough from the wall the PEs are randomly oriented.

%%%%%%%%%%%%%%%%%%%%%%%%%%%%%%%%%%%%%
\subsubsection{Fluid charge
 \label{Sec.fludi_charge_Qp-64}}
%%%%%%%%%%%%%%%%%%%%%%%%%%%%%%%%%%%%%

\begin{figure}
\includegraphics[width = 8.0 cm]{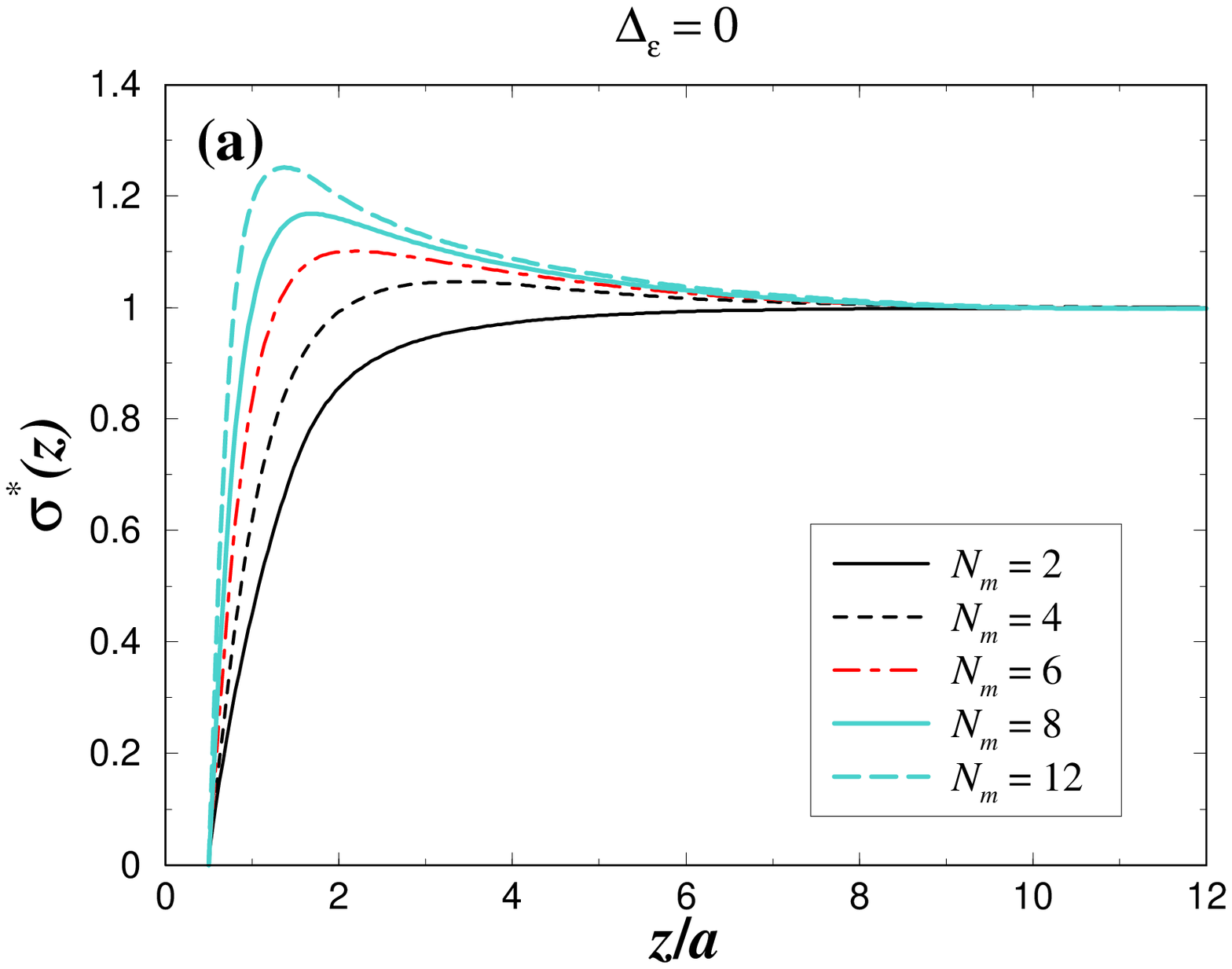}
\includegraphics[width = 8.0 cm]{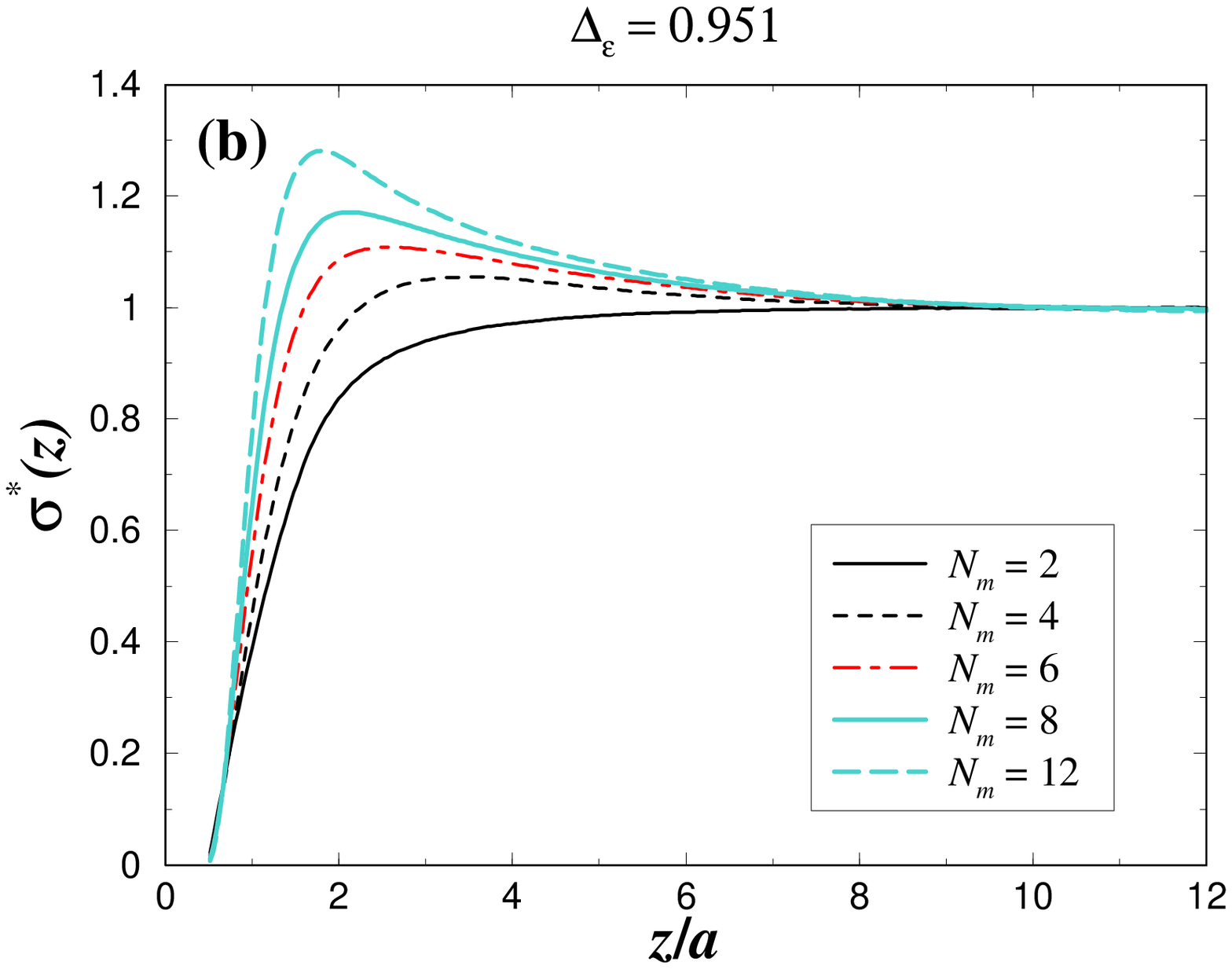}
\caption{
Profiles of the reduced net fluid charge $\sigma^*(z)$ 
for different chain length $N_m$ with $\sigma_0L^2=64$ (systems $A-E$).
(a) $\Delta_{\epsilon}=0$.
(b) $\Delta_{\epsilon}=0.951$.
}
\label{fig.Qz_star_Qp-64}
\end{figure}

Another interesting property is provided by the net fluid charge parameter 
$\sigma^*(z)$ [Eq.~\eqref{Eq.Qz_star}] that describes the screening 
of the charged interface.
The profiles of $\sigma^*(z)$ for different $N_m$ can be found in 
Fig. \ref{fig.Qz_star_Qp-64}.
At $\Delta_{\epsilon}=0$ [see Fig. \ref{fig.Qz_star_Qp-64}(a)], it is shown that
for long enough chains (here $N_m \geq 4$) the substrate gets 
locally \textit{overcharged} as signaled by $\sigma^*(z)>1$. 
Physically, this means that the (integrated) local charge of the adsorbed monomers 
\cite{note_overscreening} is larger in absolute value than that of the substrate's surface charge. 
In other words, the plate is \textit{overscreened} by the adsorbed PE chains. 
Fig. \ref{fig.Qz_star_Qp-64}(a) indicates that the degree of overcharging increases 
with $N_m$ as expected from the behavior of multivalent counterions. 
Upon inducing polarization charges 
[$\Delta_{\epsilon}=0.951$ - see Fig. \ref{fig.Qz_star_Qp-64}(b)]
overscreening is {\it maintained} and weakly disturbed, proving that the 
latter is robust against repulsive image forces.

%%%%%%%%%%%%%%%%%%%%%%%%%%%%%%%%%%%%%%%%
% FIG 6
%%%%%%%%%%%%%%%%%%%%%%%%%%%%%%%%%%%%%%%%
%

%%%%%%%%%%%%%%%%%%%%%%%%%%%%%%%%%%%%%%%%%%
\subsection{Influence of substrate's surface-charge density
 \label{Sec.charge_density}}
%%%%%%%%%%%%%%%%%%%%%%%%%%%%%%%%%%%%%%%%%%

To complete our investigation, we would like to address the
influence of the substrate charge density on the PE
adsorption.
In this respect, we consider (at fixed $N_m=8$) two additional 
values for the surface charge density:
$\sigma_0L^2=32 ~ {\rm and} ~ 128$ corresponding to the systems
$F$ and $G$, respectively (see Table \ref{tab.simu-runs}).

%%%%%%%%%%%%%%%%%%%%%%%%%%%%%%%%%%%%%
\subsubsection{Monomer distribution
 \label{Sec.mon_dist_Nm-8}}
%%%%%%%%%%%%%%%%%%%%%%%%%%%%%%%%%%%%%

%%%%%%%%%%%%%%%%%%%%%%%%%%%%%%%%%%%%%%%%
% FIG 7
\begin{figure}
\includegraphics[width = 8.0 cm]{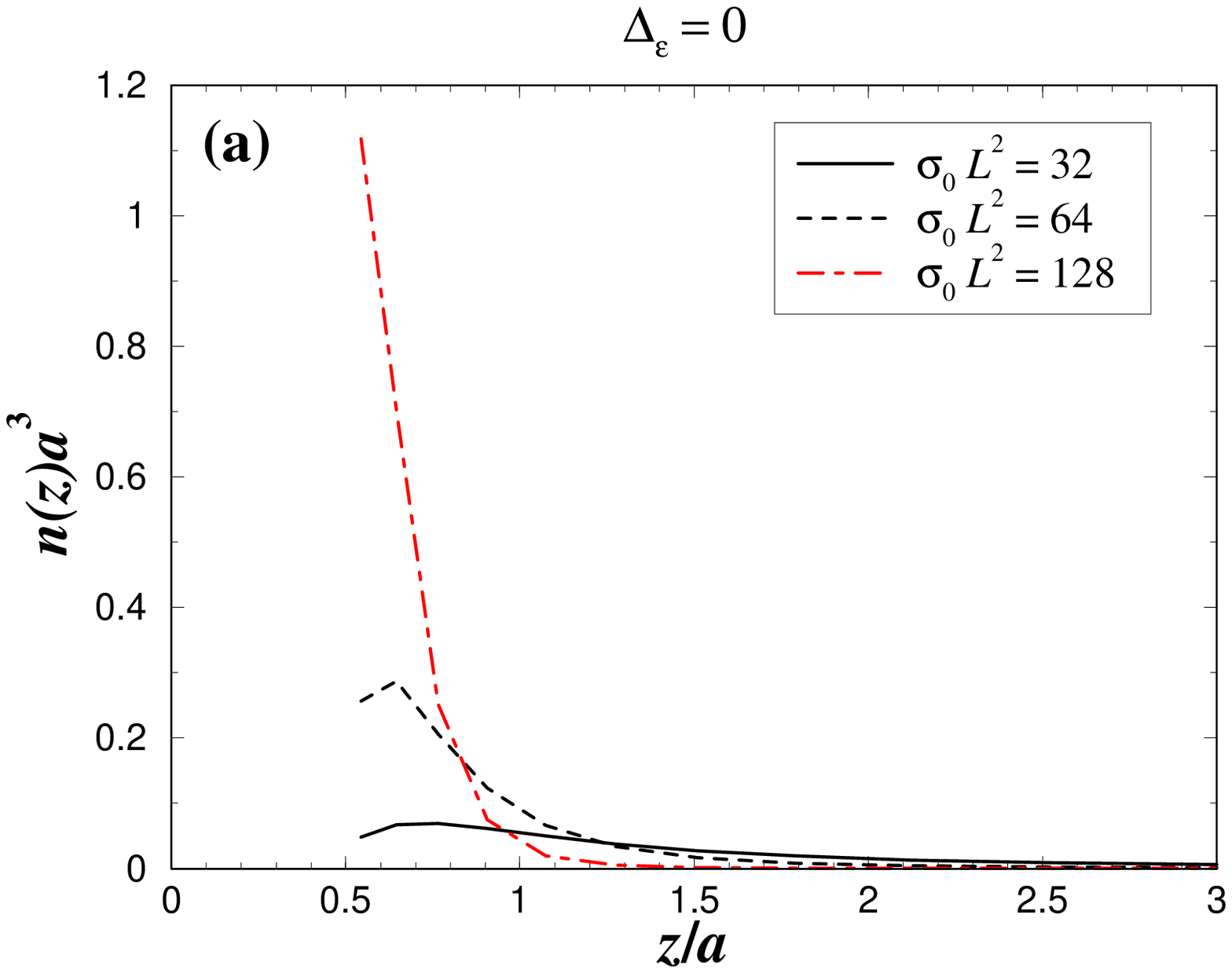}
\includegraphics[width = 8.0 cm]{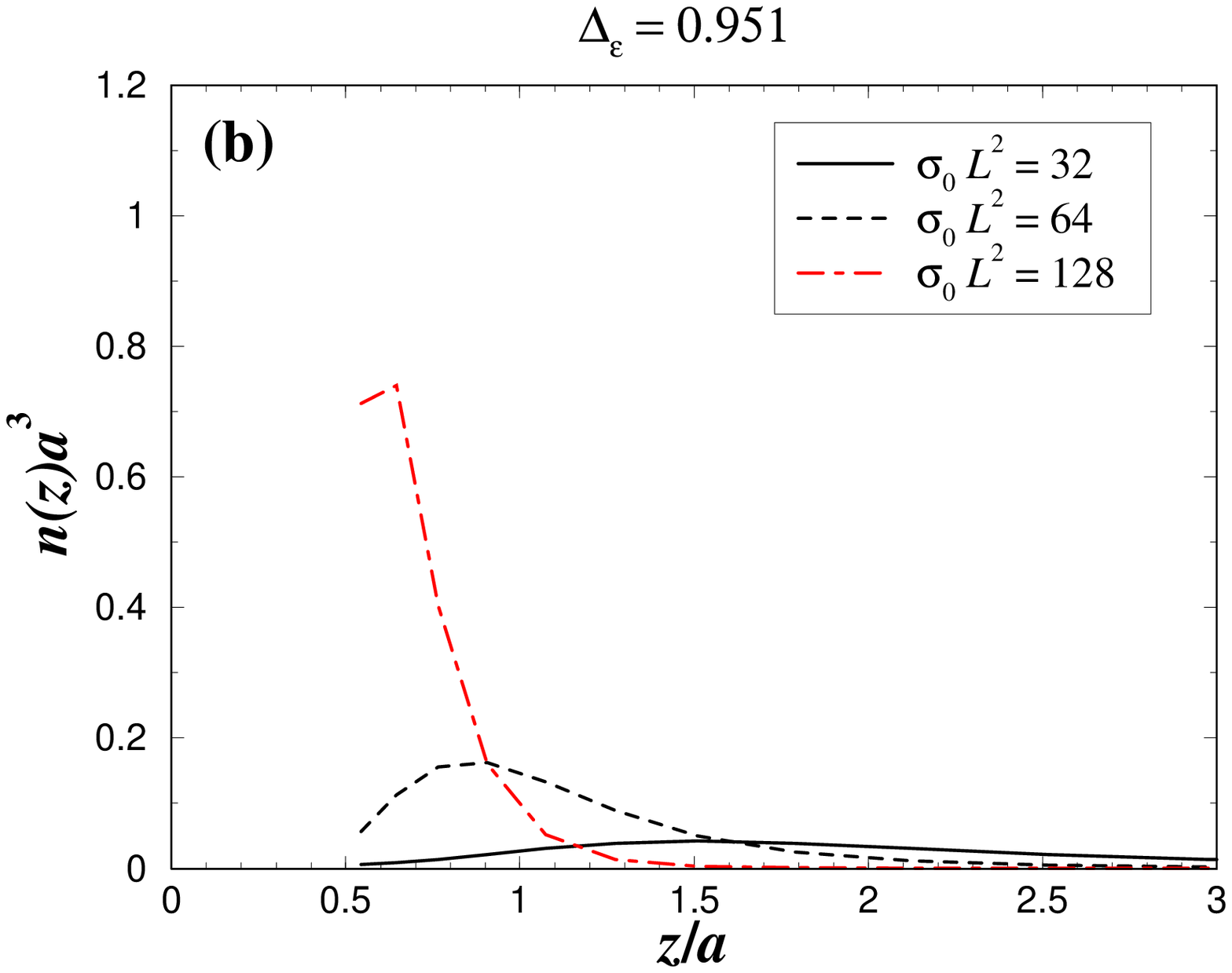}
\caption{
Profiles of the monomer density $n(z)$ for different parameters of surface-charge 
density $\sigma_0L^2$ with $N_m=8$ (systems $D,F,G$). 
The case $\sigma_0L^2=64$ (system $D$) from Fig. \ref{fig.nz_Qp-64} 
is reported here again for easier comparison.
(a) $\Delta_{\epsilon}=0$.
(b) $\Delta_{\epsilon}=0.951$. 
}
\label{fig.nz_Nm-8}
\end{figure}
%%%%%%%%%%%%%%%%%%%%%%%%%%%%%%%%%%%%%%%%
%

%%%%%%%%%%%%%%%%%%%%%%%%%%%%%%%%%%%%%%%%
% FIG 8
\begin{figure}
\includegraphics[width = 8.0 cm]{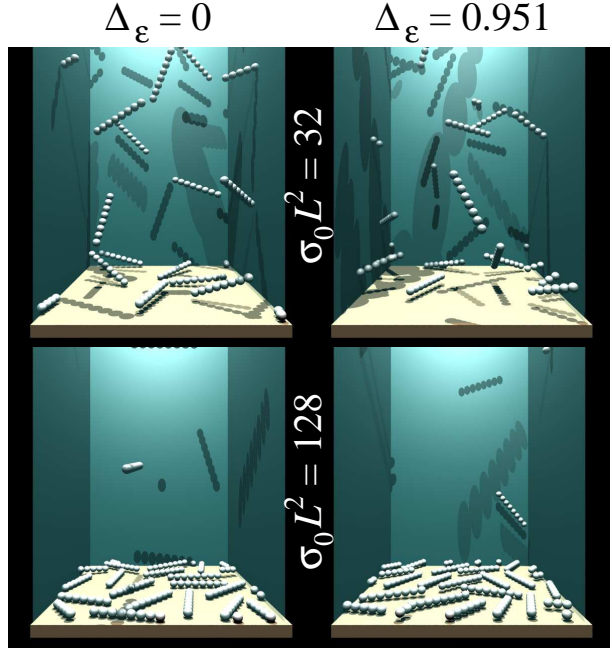}
\caption{
Typical equilibrium microstructures of systems $F$ and $G$.
The little counterions are omitted for clarity.
}
\label{fig.snap_Nm-8}
\end{figure}
%%%%%%%%%%%%%%%%%%%%%%%%%%%%%%%%%%%%%%%%
%

The plots of the monomer density $n(z)$ for various values 
of $\sigma_0L^2$ can be found in Fig. \ref{fig.nz_Nm-8}.
Typical microstructures of systems $F$ and $G$ are sketched in
Fig. \ref{fig.snap_Nm-8}.
At $\Delta_{\epsilon}=0$ [see Fig. \ref{fig.nz_Nm-8}(a)],
the monomer density at contact increases with $\sigma_0$ as it should be.
Interestingly, the local maximum in $n(z)$, present at small 
$\sigma_0$ (here $\sigma_0 L^2 \leq 64$), \textit{vanishes at large} $\sigma_0$
[see Fig. \ref{fig.nz_Nm-8}(a)].
This feature is the result of a $\sigma_0$-enhanced driving force of 
adsorption that overcomes chain-entropy effects at large enough $\sigma_0$.
The strong adsorption at $\sigma_0 L^2 = 128$ leads to a \textit{flat}
PE layer as well illustrated in Fig. \ref{fig.snap_Nm-8}.

By polarizing the interface ($\Delta_{\epsilon}=0.951$), it can be
seen from Fig. \ref{fig.nz_Nm-8}(b) and the snapshot from Fig. \ref{fig.snap_Nm-8}
that there is a strong monomer depletion near the interface for $\sigma_0 L^2 = 32$. 
This feature is due to the combined effects of %(i) PE configurational entropy, 
(i) image-PE repulsion and (ii) a lower electrostatic wall-PE attraction.
Upon increasing $\sigma_0$ the monomer density near contact becomes larger, 
and concomitantly, the location of the maximum in $n(z)$ is systematically 
shifted to smaller $z$.
It is to say that the thickness of the adsorbed PE layer decreases with $\sigma_0$.
We also expect that, at very large $\sigma_0$ (not reported here), 
this maximum vanishes leading to a purely attractive effective wall-PE interaction.

%%%%%%%%%%%%%%%%%%%%%%%%%%%%%%%%%%%%%
\subsubsection{PE orientation
 \label{Sec.legendre_Nm-8}}
%%%%%%%%%%%%%%%%%%%%%%%%%%%%%%%%%%%%%

%%%%%%%%%%%%%%%%%%%%%%%%%%%%%%%%%%%%%%%%
% FIG 9
\begin{figure}
\includegraphics[width = 8.0 cm]{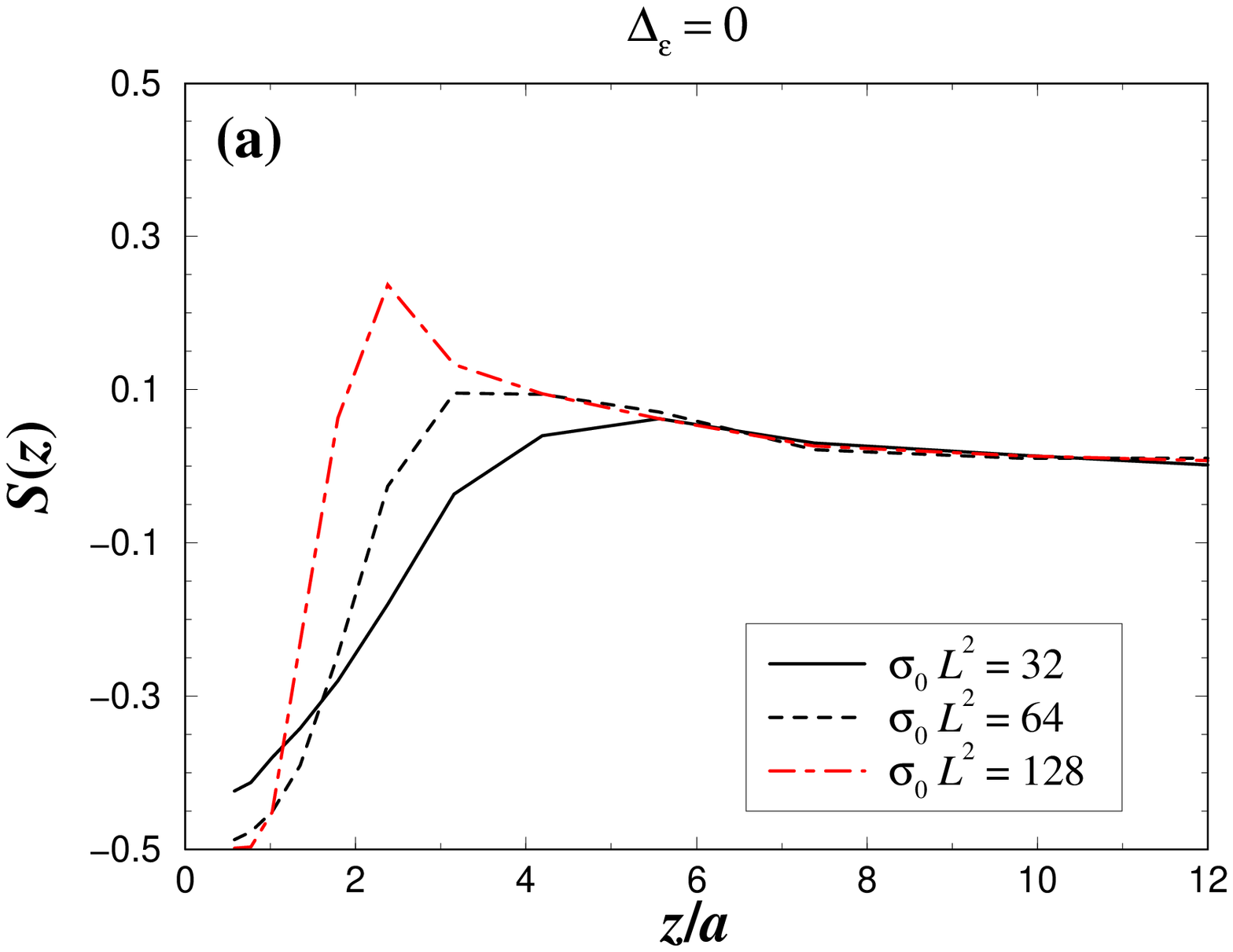}
\includegraphics[width = 8.0 cm]{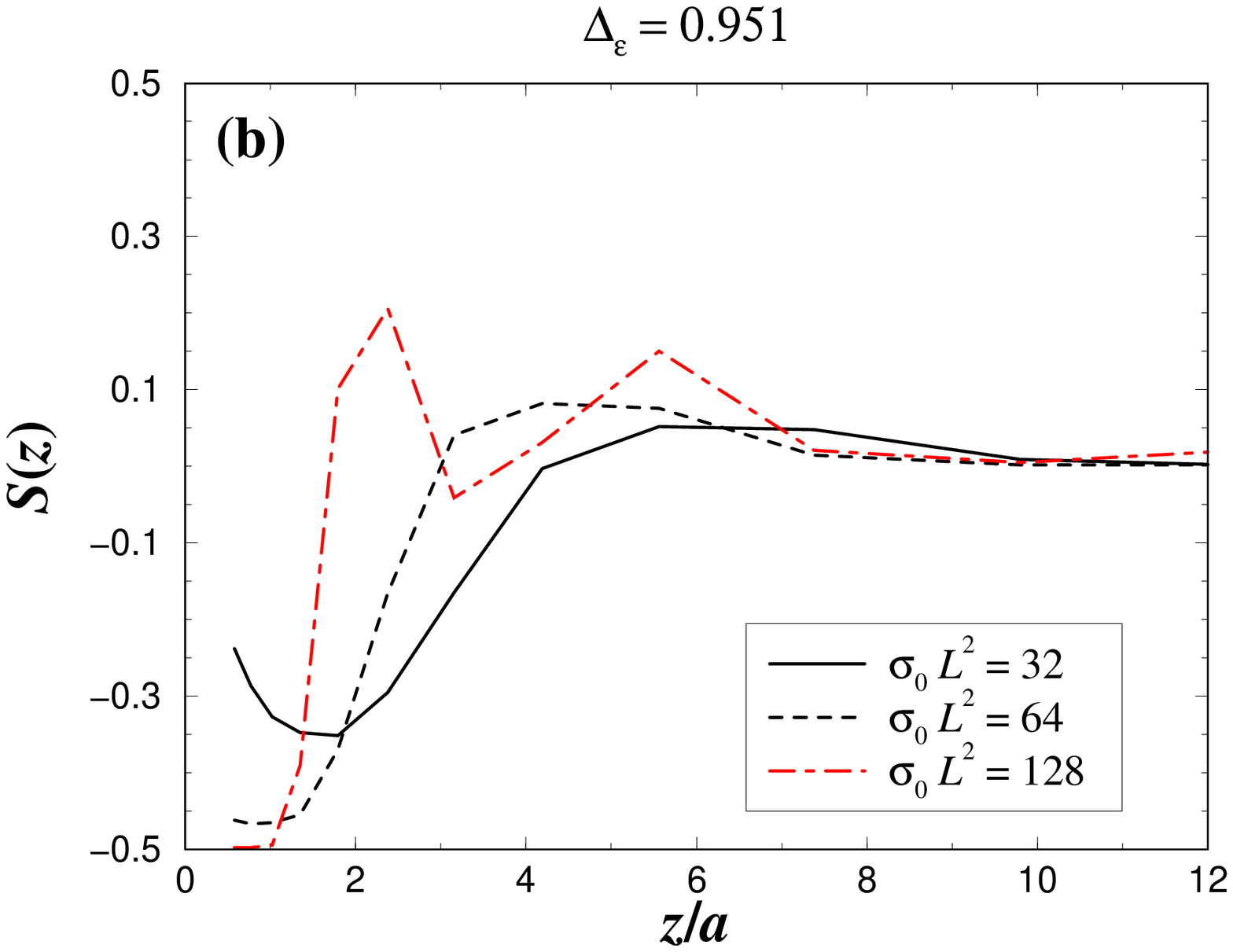}
\caption{
Same as Fig. \ref{fig.nz_Nm-8} but for $S(z)$.
}
\label{fig.Sz_Nm-8}
\end{figure}
%%%%%%%%%%%%%%%%%%%%%%%%%%%%%%%%%%%%%%%%
%

At $\Delta_{\epsilon}=0$, Fig. \ref{fig.Sz_Nm-8}(a) shows 
that near the charged interface (when $z \lesssim a$)
the degree of parallelism between the rod-like PE and the 
interface increases with growing $\sigma_0$ as indicated by
$S(z) \to -1/2$. This observation is merely due to the electrostatic 
wall-PE binding whose strength scales like $\sigma_0$ at
fixed $N_m$.

In the presence of image forces ($\Delta_{\epsilon}=0.951$),
Fig. \ref{fig.Sz_Nm-8}(b) demonstrates again for $\sigma_0L^2=32$
[see also Fig. \ref{fig.Sz_Qp-64}(b) for comparison)]
a strongly non-monotonic behavior of $S(z)$ near the interface.
%In the vicinity of the interface, it is found that
%the rod-like PEs tend to get less parallel to the interface-plane
%as $\sigma_0$ grows. 
This feature is fully consistent with the ideas that (i)
image forces become especially strong at low $\sigma_0$ and (ii)
repulsive image forces induce orientational disorder as
previously established.
This finding leads to the important general statement:
%
% IMAGE FORCES INDUCE DISORDER
%%%%%%%%%%%%%%%
\begin{itemize}
\item \textit{Repulsive} image forces at low surface charge density 
induce \textit{orientational disorder} near the interface.
\end{itemize}
%%%%%%%%%%%%%%%

%%%%%%%%%%%%%%%%%%%%%%%%%%%%%%%%%%%%%
\subsubsection{Fluid charge
 \label{Sec.fludi_charge_Nm-8}}
%%%%%%%%%%%%%%%%%%%%%%%%%%%%%%%%%%%%%

\begin{figure}
\includegraphics[width = 8.0 cm]{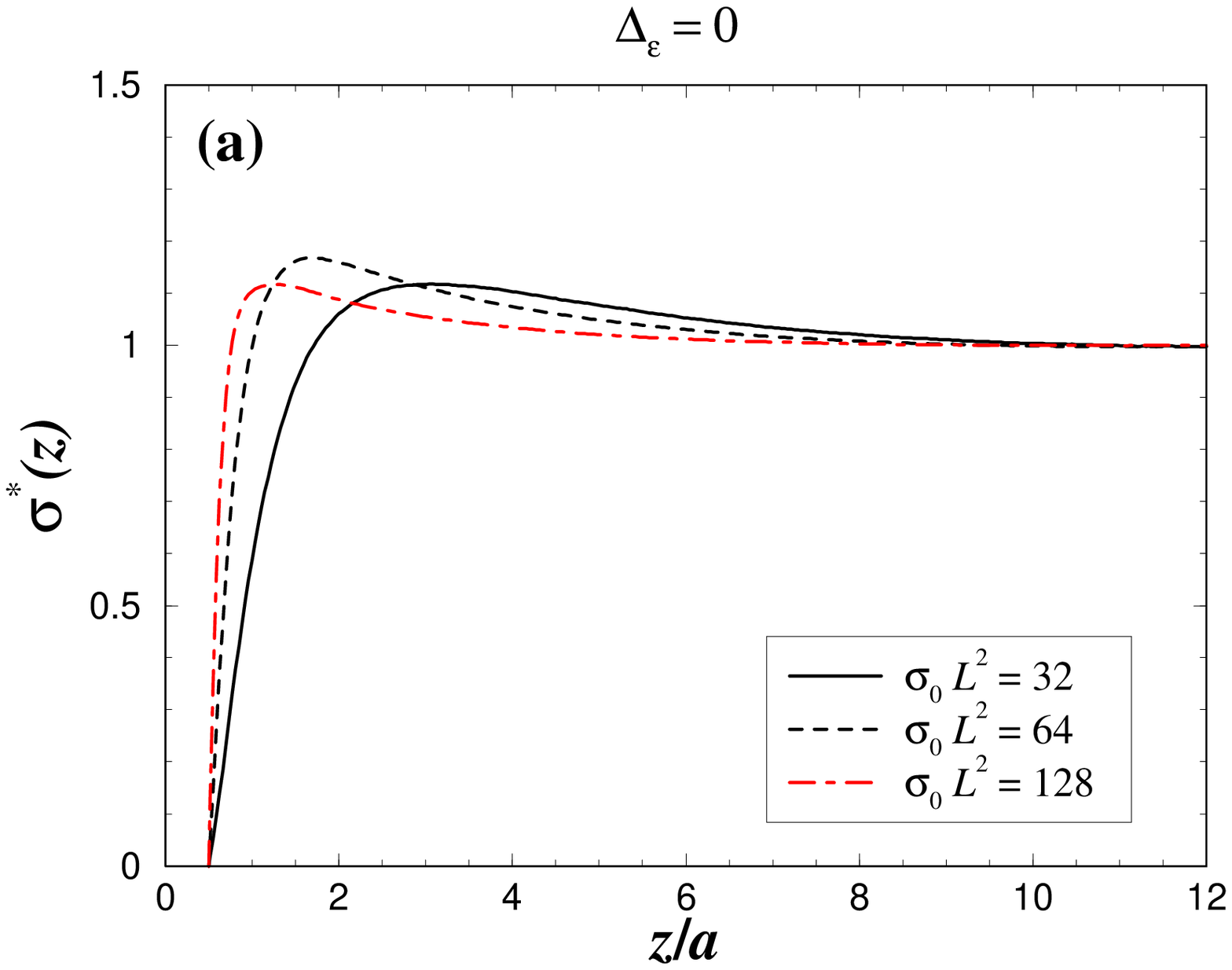}
\includegraphics[width = 8.0 cm]{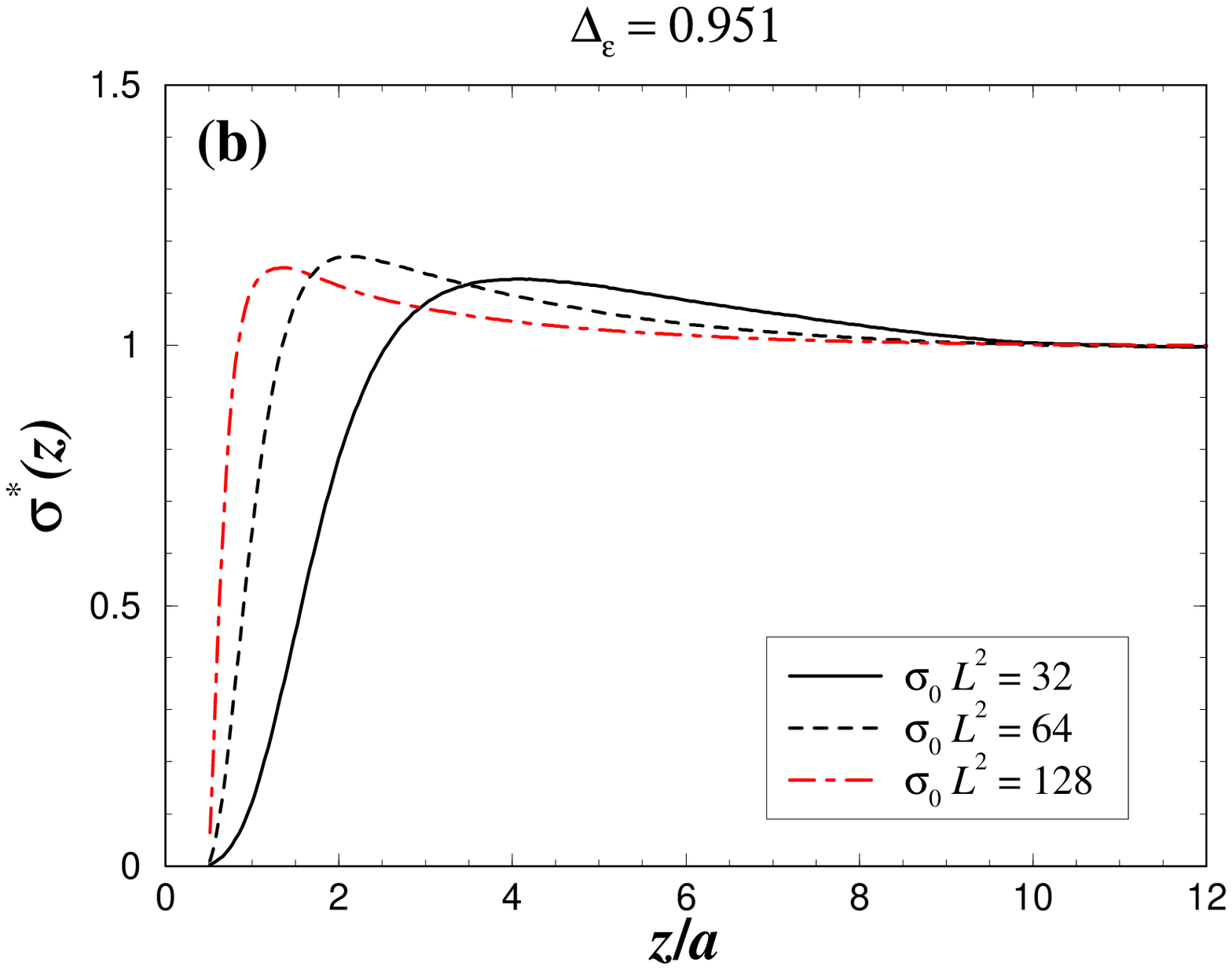}
\caption{
Same as Fig. \ref{fig.nz_Nm-8} but for $\sigma^*(z)$.
}
\label{fig.Qz_star_Nm-8}
\end{figure}

The profiles of $\sigma^*(z)$ for different $\sigma_0L^2$ can be found in 
Fig. \ref{fig.Qz_star_Nm-8}.
At $\Delta_{\epsilon}=0$ [see Fig. \ref{fig.Qz_star_Nm-8}(a)], it is found that
the planar interface gets always locally overcharged as signaled by $\sigma^*(z)>1$. 
The location of the maximum in $\sigma^*(z)$ is shifted to lower $z$ with 
increasing $\sigma_0$.

Upon inducing polarization charges 
[$\Delta_{\epsilon}=0.951$ - see Fig. \ref{fig.Qz_star_Nm-8}(b)]
overscreening is still there. 
However, at $\sigma_0L^2=64$, there is a non-negligible shift of the maximum
of about $0.5a$. The distance at which the substrate is compensated 
[i.e., where $\sigma^*(z)=1$] corresponds to 1.72a (2.54a) for
$\Delta_{\epsilon}=0$ ($\Delta_{\epsilon}=0.951$) leading to a 
neutralization $z$-shift of $0.72a$.
 
%%%%%%%%%%%%%%%%%%%%%%%%%%%%%%%%%%%%%%%%
% FIG 10
%%%%%%%%%%%%%%%%%%%%%%%%%%%%%%%%%%%%%%%%
%

%%%%%%%%%%%%%%%%%%%%%%%%%%%%%%%%%%%%%
\section{Summary
 \label{Sec.summary}}
%%%%%%%%%%%%%%%%%%%%%%%%%%%%%%%%%%%%%

To conclude, we have performed MC simulations to address the behavior of
rod-like polyelectrolytes at oppositely charged planar surfaces.
The influence of image forces stemming from the dielectric discontinuity
at the charged interface was also analyzed.
We have considered a finite and \textit{fixed} monomer concentration in the 
dilute regime for relatively short chains.

In the absence of image forces ($\Delta_{\epsilon} = 0$), 
our main findings can be summarized as follows:
%%%%%%%%%%%%%%%
\begin{itemize}
% n(m)
\item 
At moderately charged interfaces, only (\textit{very}) short rod-like PEs 
(here $2 < N_m \leq 8$) experience a short-ranged 
repulsion near the interface. For longer rod-like PEs the effective 
wall-PE interaction becomes purely attractive.
This behavior is in contrast to that occurring  with \textit{flexible} PEs,
where the chain-entropy is larger leading to stronger entropy-driven depletion. 
% S(z)
\item 
Near the charged interface, the rod-like PEs 
get more and more parallel to the interface-plane when
the chain length $N_m$ is increased.
Concomitantly, the strength of the substrate-charge reversal is
$N_m$-enhanced.
% \sigma_0
\item
Upon increasing the substrate-surface-charge density $\sigma_0$ 
it was demonstrated that:
(i) The monomer adsorption is enlarged and the wall-PE effective interaction becomes purely
repulsive for high enough $\sigma_0$.
(ii) The degree of parallelism (near the interface) between the interface-plane
and the rod-like PE is enhanced.
\end{itemize}

% Image forces
The main effects stemming from repulsive image forces 
($\Delta_{\epsilon} = 0.951$) as revealed by this study are as follows:
\begin{itemize}
% n(z)
\item
The monomer adsorption is reduced at finite $\Delta_{\epsilon}$
and the $n(z)$-profiles become  similar to those obtained with 
\textit{flexible} chains, in contrast to what was reported at 
$\Delta_{\epsilon} = 0$.
% S(z)
\item
Repulsive image forces induce PE orientational disorder near the interface.
% Q
\item
The substrate-charge reversal is robust against repulsive 
image forces.

\end{itemize}

%A future work will address the adsorption of \textit{rod}-like PEs.
%This very interesting situation was recently theoretically examined 
%by Cheng and de la Cruz \cite{Cheng_JCP_2003}. 
%Numerical simulation data would be of great help to further characterize the transversal and 
%in-plane structures as well as to elucidate the influence of image forces.  

%%%%%%%%%%%%%%%%%%%%%%%%%%%%%%%%%%%%%
%Acknowledgments
%%%%%%%%%%%%%%%%%%%%%%%%%%%%%%%%%%%%%
\acknowledgments 
R. M. thanks H. L\"owen and R. Blaak for fruitful discussions.
Financial support from DFG via LO418/12 and SFB TR6 is also acknowledged.
%The author thanks F. Caruso, H. L\"owen, S. K. Mayya and E. P\'erez for 
%helpful and stimulating discussions.  
%This work was supported by \textit{Laboratoires Europ\'eens Associ\'es} (LEA) and the SFB 625.

%\bibliographystyle{prsty}
%\bibliography{pe}

\begin{thebibliography}{10}
\bibitem{note_overscreening}
Near the charged wall, the presence of the counterions at $\Delta_{\epsilon}=0$
  is marginal so that the fluid charge is quasi exclusively provided by the
  charges of the adsorbed monomers. This becomes especially vivid when $N_m
  \geq 4$ where overcharging occurs.
\end{thebibliography}

\end{document}